\documentclass[prd,twocolumn,reprint,preprintnumbers,nofootinbib,superscriptaddress]{revtex4-2}

\input{header.tex}

\usepackage{xcolor}
\newif\ifshowchanges
\ifshowchanges
  \newcommand{\ch}[1]{{\color{blue}#1}}
\else
  \newcommand{\ch}[1]{#1}
\fi

\begin{document}

\reportnum{-2}{CERN-TH-2026-130}

\title{Detector performance at SHiP for cascade-produced long-lived particles}
\author{Matei~Climescu}
\email{matei.climescu@cern.ch}
\affiliation{Ghent University, Ghent, Belgium}
\author{Yehor~Kyselov}
\email{kiselev883@gmail.com}
\affiliation{Taras Shevchenko National University of Kyiv, Kyiv, Ukraine}
\author{Maksym~Ovchynnikov}
\email{maksym.ovchynnikov@cern.ch}
\affiliation{Theoretical Physics Department, CERN, 1211 Geneva 23, Switzerland}

\date{\today}

\begin{abstract}
Previous studies have shown that cascade production in the thick target of the SHiP experiment may substantially enhance the number of light long-lived particles (LLPs) decaying in the fiducial volume. However, cascade-produced LLPs are typically soft, so daughter-level acceptance and reconstruction effects can strongly suppress the observable event rate. We quantify this suppression for two representative cases: photophilic axion-like particles (ALPs) produced in electromagnetic cascades, and heavy neutral leptons (HNLs) produced in decays of secondary kaons. We combine a semi-analytic event-rate calculation with a detector-level study of ALP reconstruction in the electromagnetic calorimeter. For the nominal SHiP detector design, cascade ALPs give at most a moderate enhancement over primary production, and only at the lightest masses; at higher masses, the cascade contribution becomes subdominant or negligible. For HNLs from secondary kaons, the cascade contribution is already subdominant after imposing daughter-level geometric acceptance. We also identify possible ways to recover part of the cascade event rate, including relaxed event-selection criteria and an active decay-volume subdetector.
\end{abstract}

\maketitle

\section{Introduction}
\label{sec:introduction}

Hypothetical particles with GeV-scale masses and tiny couplings to Standard Model (SM) fields may address several open problems in particle physics~\cite{Asaka:2005an,Shaposhnikov:2006nn,Pospelov:2007mp,Bezrukov:2009yw,Krnjaic:2015mbs,Berlin:2018jbm,Hook:2019qoh}. Despite their small couplings, such particles can be efficiently produced at accelerator experiments operating at much higher energy scales~\cite{Gorbunov:2007ak,Ilten:2018crw,Aloni:2018vki,Bondarenko:2018ptm,Berlin:2018jbm,Boiarska:2019jym,Foroughi-Abari:2021zbm,Ovchynnikov:2025gpx}, and can be searched for using a wide range of techniques and signatures~\cite{Alekhin:2015byh,Ilten:2016tkc,Feng:2017uoz,Belle-II:2018jsg,Beacham:2019nyx,NA64:2020qwq,Blondel:2022qqo,Antel:2023hkf,Gorkavenko:2023nbk,DUNE:2024wvj,deBlas:2025gyz,Balkin:2025rtc}. They are often referred to as long-lived particles, or LLPs.

The upcoming SHiP experiment at SPS~\cite{Aberle:2839677,SHiP:2025ows}, to be built at the SPS North Area, is one of the flagships in exploring the parameter space of GeV-mass LLPs via the signatures of displaced decays and scatterings, for a broad range of LLP scenarios~\cite{Beacham:2019nyx,Antel:2023hkf,deBlas:2025gyz}. This is achieved through a very high intensity of the SPS proton beam with the relatively large collision energy, together with excellent background suppression and close-to-optimal detector geometry~\cite{Bondarenko:2023fex}.

SHiP is currently in the TDR phase~\cite{Albanese:2948477}, making detector-performance studies increasingly important for realistic LLP signals and kinematic regimes. Most SHiP sensitivity studies~\cite{SHiP:2018xqw,SHiP:2020vbd,Aberle:2839677} assume that LLPs are produced predominantly in the first interaction of the proton beam with the target. However, several recent works~\cite{SHiP:2020noy,Gorbunov:2020rjx,Blinov:2024pza,Zhou:2024aeu,Blinov:2025aha,Patrone:2025fwk} have shown that secondary interactions inside the thick target can yield additional LLPs. These cascade-produced LLPs are typically much softer than those from primary production. It is therefore important to assess whether their decay products remain within the detector acceptance and can be reconstructed with sufficient efficiency.

In this work, we study the SHiP detector response to such low-energy cascade-induced signals. We extend previous estimates in two steps. First, using a semi-analytic calculation, we evaluate the daughter-level geometric acceptance, namely the fraction of LLP decays for which the visible decay products reach the relevant downstream detector. Second, for the two-photon final state, we perform a detector-level simulation and reconstruction study for events passing this geometric requirement.

We consider two representative scenarios: axion-like particles (ALPs) coupled to photons, which can be efficiently produced in electromagnetic cascades, and heavy neutral leptons (HNLs), which may be produced in decays of secondary kaons. The geometric-acceptance part of our analysis can be generalized to other low-energy signals, including vector particles such as dark photons or mediators coupled to the $L_{\alpha}-L_{\beta}$ current~\cite{Zhou:2024aeu,Blinov:2025aha}.

The paper is organized as follows. Sec.~\ref{sec:ship} describes the SHiP experiment and the baseline selection criteria. Sec.~\ref{sec:cascades} reviews the main sources of cascade-produced LLPs in the thick SHiP target. Sec.~\ref{sec:existing-studies} summarizes existing estimates and their limitations for low-energy cascade signals. Sec.~\ref{sec:our-approach} presents our calculation of the LLP fluxes, geometric acceptance, and reconstruction efficiency. Sec.~\ref{sec:results} contains the main results and discussion: for ALPs, we compare the semi-analytic acceptance estimate with the detector-level reconstruction study, quantifying the impact of the ECAL acceptance, clustering, and impact parameter requirements, and discuss possible ways to improve the sensitivity to low-energy two-photon signals; for HNLs, we show that the cascade contribution becomes subdominant once daughter-level geometric acceptance is imposed. We conclude in Sec.~\ref{sec:conclusions}.

\section{SHiP experiment}
\label{sec:ship}

\begin{figure}
    \centering
    \includegraphics[width=\linewidth]{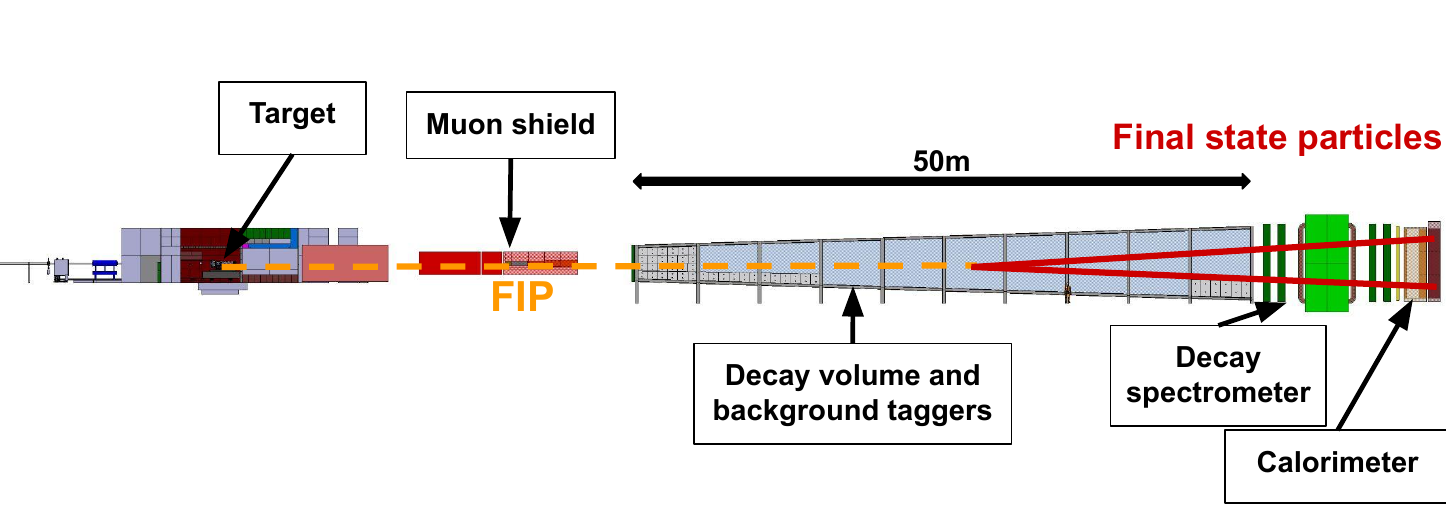}
    \caption{Representation of the SHiP detector overlayed with a 2-body FIP decay.}
    \label{fig:ship_fig}
\end{figure}

The Search for Hidden Particles (SHiP) experiment is a proposed beam-dump facility at the CERN SPS designed to search for feebly interacting particles with macroscopic lifetimes. The experiment will use a slow-extracted proton beam with momentum $400~\gev$ and an annual intensity of about
\begin{equation}
    N_{\rm PoT} \simeq 4\times 10^{19}\, ,
\end{equation}
directed onto a compact high-$Z$ target. A representation of the experiment is shown in Fig.~\ref{fig:ship_fig}. The target serves as both the production point for hidden-sector particles and the primary beam dump. It has a transverse size of order $30~\cm \times 30~\cm$ and a longitudinal depth of order $1.2$--$1.4\,\m$, corresponding to approximately $12$ nuclear interaction lengths. The baseline design uses a dense tungsten-based core. This choice maximizes the production of heavy mesons and photons, while suppressing pion and kaon decays in
flight, thereby reducing the neutrino and muon backgrounds entering the detector region.

Immediately downstream of the target, the experimental layout contains a hadronic absorber and a magnetic muon shield. Their purpose is to absorb secondary hadrons and to sweep away the intense flux of muons produced in
the dump before these particles can reach the hidden-sector decay volume. This shielding system is essential for achieving the low-background environment required for searches for displaced visible decays.

The hidden-sector decay volume starts at
$z \simeq 33\,\m$ and ends at $z \simeq 83\,\m$, giving a fiducial length of approximately $50\,\m$. Its geometry is approximately that of a truncated
pyramid, aligned with the beam axis. The transverse opening is about $1\,\m \times 3\,\m$ at the entrance and expands to about $4\,\m \times 6\,\m$ at the downstream end. The volume is filled with a helium balloon to minimize secondary interactions in the material while maintaining a large geometrical acceptance for long-lived particle decays.

A magnetic spectrometer composed of four straw-tube-based tracking stations is located downstream of the decay volume, after a small $\simeq 1\,\m$ gap. In the current layout, the spectrometer system occupies approximately the region
\begin{equation}
    84\,\m \lesssim z \lesssim 95\,\m \,.
\end{equation}
It consists of a large-aperture dipole magnet and four low-mass tracking stations, arranged with tracking stations before and after the magnetic-field region. The magnet has an aperture of approximately $4\,\mathrm{m}\times 6\,\mathrm{m}$ and provides a field integral of $\int B\,dl \simeq 0.6\text{--}0.8\,\mathrm{Tm}$, with a peak field of about $B\simeq 0.15\,\mathrm{T}$. The curvature of charged-particle trajectories in the magnetic field provides the momentum measurement, while the reconstructed tracks are used to determine the decay vertex, the impact parameter with respect to the production target, and the invariant mass under the relevant particle-identification hypotheses.

A timing detector is placed immediately downstream of the spectrometer, around $z \simeq 95\,\m$. It provides precise time information for charged tracks and helps reject accidental combinations of particles from uncorrelated backgrounds.

The electromagnetic calorimeter (ECAL) begins at
$z \simeq 95.5\,\m$ and ends at $z\approx 97.32\,\m$. Its transverse dimensions are approximately
\begin{equation}
    A_{\rm ECAL} \approx 4.32\,\m \times 6.48\,\m \,.
\end{equation}
It uses the SplitCal detector concept~\cite{Bonivento:2018eqn,Climescu:2025kdj}, which provides electron and photon identification, measures electromagnetic shower energies, and reconstructs shower directionality. In combination with the spectrometer, it is crucial for reconstructing final states containing electrons or photons, including two-photon signatures and partially reconstructed hidden-sector decays.

\begin{figure}
\includegraphics[width=0.45\textwidth]{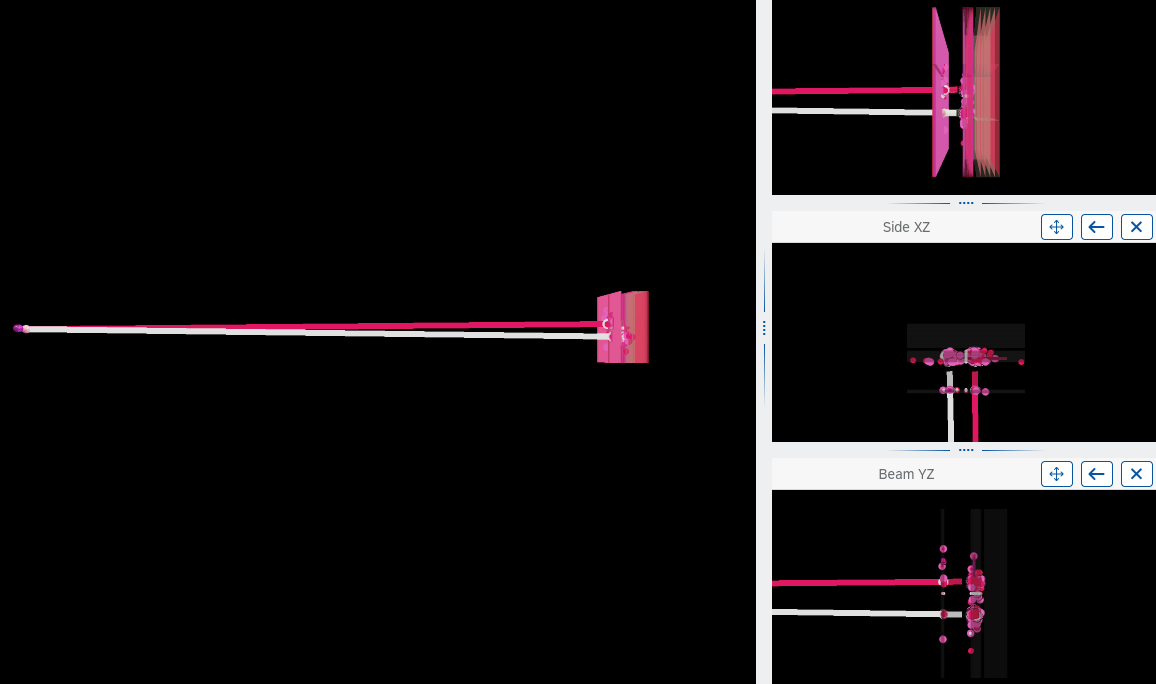}
\caption{Event display of a reconstructed \SI{10}{\giga\electronvolt} $\pi^0\rightarrow \gamma\gamma$ event. The magenta and white lines correspond to the reconstructed photon directions; the pink dot on the left panel corresponds to the reconstructed vertex.}
\label{fig:event_display}
\end{figure}

The calorimeter is composed of scintillator bars and three \textit{High Precision Layers} (HPLs) based on scintillating fibres to improve angular resolution. The full detector comprises 46 layers, with a \SI{1}{\meter} longitudinal gap separating the first 5 $X_0$ from the remainder of the detector, with the high-precision layers placed after 5 (right before the gap), 7, and 11 $X_0$. An event display is shown in Fig. \ref{fig:event_display}.  The center of the calorimeter is placed \SI{96.97}{\meter} away from the tungsten target, which is also the origin.

Finally, the hadronic calorimeter (HCAL) is located downstream of the ECAL. In the current layout, it starts at $z \simeq 97.42~\m$ and ends at $z \simeq 98.42~\m$, and it is arranged as a 5$\lambda_A$ iron sandwich calorimeter sampled every $\lambda_A$ by scintillator layers similar to the ECAL's \cite{Climescu:2025kdj}.

The HCAL provides muon and hadron identification and measures the energy deposited by hadronic showers, complementing the ECAL information in the downstream particle identification system.

The entire calorimeter system provides particle identification in excess of 99\% true reconstruction for electrons, charged pions, and muons.

\subsection{Baseline selection criteria}
\label{sec:ship-baseline-selection}

Let us summarize the selection criteria on the events with decaying LLPs used in previous SHiP studies, following Refs.~\cite{Aberle:2839677,SHiP:2025ows}. 

The central signature is a displaced decay vertex reconstructed from the visible decay products. The baseline requirement is at least two reconstructible visible objects with total electric charge zero, such as $\gamma\gamma$, $e^{+}e^{-}, \mu^{\pm}\pi^{\mp}$, or other states.

Charged decay products are required to be reconstructed in the magnetic tracking spectrometer, which provides their momenta, the decay vertex, and the main kinematic information for charged final states. In the reconstruction considered here, charged tracks are required to lie within the acceptance of all four tracking stations of the spectrometer. The downstream particle-identification system, consisting primarily of the ECAL and HCAL, is then used to assign particle hypotheses and to reconstruct neutral electromagnetic final states such as photons. Thus, the spectrometer alone can reconstruct charged-track kinematics under assumed mass hypotheses, but calorimetry is needed to distinguish electrons, hadrons, and penetrating muon-like tracks, and to reconstruct final states containing photons or neutral hadrons. This particle-identification (PID) information, together with timing, vetoes, and topological requirements, is important for SHiP to operate in a nearly background-free regime. In particular, relaxing the PID requirements, for instance by requiring electron decay products to be reconstructed only in the spectrometer without an ECAL match, would weaken background rejection. For this reason, when quantifying the geometric component of the decay-product acceptance, we require the visible decay products to remain within the relevant downstream detector acceptance: the ECAL for electromagnetic final states and the HCAL for hadronic or penetrating muon-like final states.

The SHiP collaboration studies summarize baseline selection criteria that the decay products must pass to be considered as a reconstructed decay event:
\begin{itemize}
\item Each visible decay product must have momentum, or energy in the case of photons,
\begin{equation}
p_{\text{daughter}}>1\,\gev\,,
\label{eq:selection-ship-baseline-1}
\end{equation}
where for photons this condition should be understood as $E_{\gamma}>1\,\gev$.

\item The reconstructed tracks or directions must be compatible with a common displaced vertex, quantified by a distance of closest approach (DOCA),
\begin{equation}
\text{DOCA}<1\,\cm\,.
\label{eq:selection-ship-baseline-2}
\end{equation}

\item The four-momentum of the LLP candidate, reconstructed from the decay-product kinematics, is extrapolated back toward the target; the corresponding transverse impact parameter (IP) must satisfy
\begin{equation}
\mathrm{IP}<2.5\,\m\,.
\label{eq:selection-ship-baseline-3}
\end{equation}
For fully reconstructible final states made of charged particles (all decay products are reconstructed), this requirement can be tightened to $\mathrm{IP}<0.1\,\m$.
\end{itemize}

After applying these selection criteria and requiring full event reconstruction, the search for displaced decays is expected to be effectively background-free. For the typical kinematics of LLPs $X$ produced in primary, non-cascade proton beam collisions, with energies $E_{X}\gtrsim 20\,\gev$~\cite{Ovchynnikov:2023cry}, the criteria~\eqref{eq:selection-ship-baseline-1}--\eqref{eq:selection-ship-baseline-3} generically lead to an $\mathcal{O}(1)$ event reconstruction efficiency, assuming that the LLPs mostly decay into charged states~\cite{SHiP:2018xqw,SHiP:2020vbd}. See Sec.~\ref{sec:cascades} for the distinction between primary and cascade production.

Performance of the SHiP detector to reconstruct the events with decays into a pair of photons has been studied in a recent work~\cite{Climescu:2025kdj}, considering photophilic ALPs produced in primary interactions as a concrete example. For decays into a pair of photons, an additional effective requirement is that the two photons be separated in the transverse plane at the ECAL entrance, $|\Delta \mathbf{r}_{\perp}|$ by approximately one Moli\`ere radius,
\begin{equation}
|\Delta \mathbf{r}_{\perp}| >r_{\rm M}\approx 3.2\,\text{cm}\,.
\label{eq:selection-ship-baseline-4}
\end{equation}
The study confirmed good reconstruction efficiency for events satisfying the selection~\eqref{eq:selection-ship-baseline-1}--\eqref{eq:selection-ship-baseline-4}. Based on its framework, we will study how the reconstruction efficiency behaves when relaxing the energy criterion~\eqref{eq:selection-ship-baseline-1} and the impact-parameter criterion~\eqref{eq:selection-ship-baseline-3} (imposed in a tightened form in the detector-level study, see Sec.~\ref{ssec:reconstruction}), which is important for secondary LLPs produced in cascade interactions (see Sec.~\ref{ssec:selection-cuts}). As we will show there, relaxing the energy threshold is useful only in combination with a relaxed impact-parameter requirement: on its own it recovers very little signal, because the soft events it admits are in any case removed by
the impact-parameter cut.

In this work, we use these cuts as an illustration of how the event rate depends on the selection criteria used by the collaboration (see Sec.~\ref{sec:discussion-semi-analytic}), while the detector-level study below also considers modified reconstruction requirements. In particular, we avoid using a hard cut on the photon's energy. Instead, the detector-level study derives the efficiency from the simulated ECAL response, including cluster formation and reconstruction effects at low photon energies; the background implications of this choice require a dedicated study. Still, as we will see in Sec.~\ref{sec:results}, the calorimeter reconstruction efficiency typically decreases for lower-energy final states because the relative energy and angular resolutions worsen.

\section{Particles from low-energy cascades}
\label{sec:cascades}

The thick target of beam-dump experiments like SHiP not only absorbs the primary proton beam but also initiates an extended hadronic and electromagnetic cascade. Throughout this work, we distinguish between the production of LLPs via primary and cascade particles. We call an LLP primary if it is produced in the first proton-target interaction, or in the prompt interaction (decay or scattering) of particles produced in that interaction. We call an LLP cascade-produced, or secondary, if the Standard Model particle responsible for its production is itself produced later inside the shower.

Cascade production is therefore not a single mechanism, but a collection of physically distinct source classes. One important class is electromagnetic cascades, composed mainly of photons, electrons, and positrons. These particles may produce LLPs through Primakoff scattering, photon fusion, electron or photon bremsstrahlung, or related electromagnetic processes. The relevant LLP examples include photophilic ALPs, dark photons, and vector mediators coupled to lepton currents~\cite{Blinov:2024pza,Zhou:2024aeu,Blinov:2025aha,Patrone:2025fwk}. The defining feature of this source is that the secondary particles are typically much softer than the prompt photons from neutral-meson decays, while still remaining relatively aligned with the shower direction because electromagnetic interactions are dominated by small momentum transfers.

Another class is formed by secondary long-lived mesons, in particular charged and neutral kaons~\cite{Gorbunov:2020rjx}. These particles may decay in flight before reinteracting or being absorbed, producing LLPs through rare or weak decays. Examples include HNLs produced in leptonic and semileptonic kaon decays~\cite{Bondarenko:2018ptm}, as well as Higgs-like scalars and ALPs produced in rare kaon decays~\cite{Boiarska:2019jym,Ovchynnikov:2025gpx}. Compared to electromagnetic shower particles, secondary kaons have broader angular distributions, and the probability that they decay before absorption depends strongly on their energy. As a result, the corresponding LLP flux is typically softer and less forward than the flux produced by prompt charm or beauty hadrons.

A further class is provided by secondary short-lived mesons, such as $\pi^{0}$, $\eta$, and $\eta'$. These cascade mesons are typically softer and more broadly distributed than those produced in the primary proton-target interaction. As a result, the LLPs originating from their decays inherit a low-energy, large-angle kinematic profile. Such contributions are particularly relevant for light mediators produced in meson decays, including dark photons, generic vector mediators, and light dark matter candidates~\cite{SHiP:2020noy,Blinov:2024pza,Zhou:2024aeu,Blinov:2025aha}.

Finally, secondary interactions may also increase the number of heavy mesons, such as $D$ and $B$ mesons. In this case, the effect is more moderate: simulations of the SHiP target indicate an enhancement of order $\simeq 1.5-2$ for heavy-meson production~\cite{CERN-SHiP-NOTE-2015-009}. Moreover, the kinematic properties of the resulting LLPs are closer to those obtained from primary heavy-meson production than in the electromagnetic-cascade or secondary-kaon cases. LLP production from decays of secondary heavy mesons is therefore important for precision estimates of the total sensitivity, but it is not the main source of the low-energy reconstruction effects studied in this work.

A common feature of most cascade sources relevant here is that they can increase the number of produced LLPs while shifting the LLP spectrum toward lower energies. This may enhance the probability that an LLP decays inside the fiducial volume, since the decay probability in the long-lifetime regime scales approximately as $1/\gamma_{X}$. At the same time, the lower boost increases the opening angles of the visible decay products and makes them more vulnerable to geometric acceptance of the detector, energy threshold, deflection of charged particles by the spectrometer magnetic field, and reconstruction resolution. The observable impact of cascade production is therefore controlled by a competition between enhanced production and decay probability on the one hand, and detector-level suppression on the other; we discuss this in more detail in Sec.~\ref{ssec:existing-studies-limitations}.

\section{Existing studies}
\label{sec:existing-studies}

Several recent studies have investigated the impact of cascade production on the SHiP sensitivity to light LLPs. These studies differ in the source particles considered, in the treatment of the target cascade, and in the level at which the detector response is modeled. In this work, we focus on two representative cases that probe the low-energy regimes where detector-level effects are expected to be particularly important. Photophilic ALPs produced in electromagnetic cascades test the case where the secondary source is relatively forward but very soft. HNLs produced in kaon decays test the case where the secondary source is both soft and broadly distributed. The discussion and treatment of other cascade-induced signals, such as vector mediators produced in electromagnetic cascades~\cite{Zhou:2024aeu,Blinov:2025aha}, and Higgs-like scalars or ALPs produced in kaon decays, are conceptually similar to the strategy developed below.

\subsection{ALPs coupled to photons}
\label{ssec:ALP-pheno}

We first consider an ALP coupled dominantly to photons, typically described by the effective interaction
\begin{equation}
    \mathcal{L}_{a} = \frac{g_{a}}{4} a F_{\mu\nu}\tilde{F}^{\mu\nu}\,.
\end{equation}
Here, $a$ is the ALP field, $g_{a}$ is the ALP-photon coupling, $F_{\mu\nu}$ is the electromagnetic field-strength tensor, and $\tilde{F}^{\mu\nu}=\frac{1}{2}\epsilon^{\mu\nu\alpha\beta}F_{\alpha\beta}$ is its dual. 

At proton fixed-target experiments, such ALPs may be produced through several mechanisms. Primary production includes rare decays of neutral mesons
\begin{equation}
P \xrightarrow{\gamma+\gamma^{*}} \gamma +\gamma+ a, \quad P=\pi^{0},\eta,\eta'\,,
\end{equation}
photon fusion in proton-nucleus and proton-proton interactions, 
\begin{equation}
p+p(Z) \xrightarrow{\gamma^{*}+\gamma^{*}} a + p + p(Z)\,,
\end{equation}
and the Primakoff process,
\begin{equation}
\gamma+ Z\to a+ Z, \quad \gamma +p\to a + p\,,
\end{equation}
initiated by photons produced in the primary proton-target collision~\cite{Dobrich:2015jyk,Dobrich:2019dxc,Jerhot:2022chi,Ovchynnikov:2023cry}. Various electromagnetic production mechanisms may be initiated by secondary photons, electrons, and positrons generated inside the target.

The dominant ALP decay is $a\to 2\gamma$. The probabilities of other channels, like $a\to \gamma +l^{+}l^{-}$ and $a\to \gamma+{\rm hadrons}$, are at the level of $\mathcal{O}(10^{-2})$~\cite{EscuderoAbenza:2025tsi}, and are therefore neglected.

Ref.~\cite{Patrone:2025fwk} studied the ALP flux induced by secondary photons in the electromagnetic cascade inside the SHiP target. The electromagnetic shower was generated using \textsc{PETITE}~\cite{Blinov:2024pza}, a lightweight Monte Carlo tool designed to simulate electromagnetic cascades in thick targets and to provide fast interpolated shower fluxes for dark-sector production. In this framework, shower particles are generated once and then used as input for different weakly coupled particle production channels, avoiding a full resimulation of the cascade for every model point. The ALP flux was obtained with \textsc{ALPETITE}: channels with a dark-photon analogue were obtained by reweighting the corresponding \textsc{PETITE} samples, while Primakoff production was simulated with dedicated Monte Carlo samplers using the shower particles as input.

The resulting event rate in Ref.~\cite{Patrone:2025fwk} was computed by assigning weights for ALP production and decay inside the fiducial volume. Additionally, as a proxy of the decay products acceptance, the trajectory of the ALP has been required to be within $\theta_{a}<\sqrt{S_{\rm decvol}}/(2\times z_{\rm det}) \approx 0.029\,\text{rad}$, where $z_{\rm det}\approx 83\,\m$ is the distance from the target to the beginning of the SHiP spectrometer, and $S_{\rm decvol} \approx 4\,\m\times 6\,\m$ is the transverse area of the end of the decay volume. In addition, the energy-threshold requirement of $E_{a}>200\,\mev$ has been applied to each generated ALP candidate. 

The authors found that the decay event rate from secondary photons may exceed the rate from primary production modes by a factor of order $50$ for light ALPs, while the enhancement decreases with increasing mass and becomes of order unity around $m_{a}\sim 1\,\mathrm{GeV}$. This study, therefore, provides a concrete example of the general cascade mechanism discussed above, in which an enhanced low-energy source population competes with the detector requirements needed to observe the LLP decay.

\subsection{HNLs}
\label{ssec:HNL-pheno}

We next consider HNLs coupled to the Standard Model through mixing with active neutrinos. In a minimal gauge-invariant description, these mixings arise from Yukawa interactions of the form
\begin{equation}
    \mathcal{L}_{N} = Y_{\alpha}\bar{L}_{\alpha}\tilde{H}N + {\rm h.c.}\,.
\end{equation}
Here, $L_{\alpha}, \ \alpha=e,\mu,\tau$ are the Standard Model lepton doublets, $Y_{\alpha}$ are the Yukawa couplings, $\tilde{H}=i\sigma_2 H^*$ is the conjugated Higgs doublet, and $N$ is the HNL field. At the scales well below the electroweak transition, the HNL phenomenology at accelerator experiments is controlled by its mass $m_{N}$ and by the active-sterile mixing angles $U_{\alpha} \simeq Y_{\alpha}v/(\sqrt{2}m_{N})$, with $v$ being the Higgs VEV~\cite{Bondarenko:2018ptm}.

At SHiP, HNLs may be produced in decays of charm and beauty hadrons, for example
\begin{equation}
    D \to \ell_{\alpha} + N\,,\qquad D \to h+ \ell_{\alpha}+ N\,,
\end{equation}
and similarly
\begin{equation}
    B \to \ell_{\alpha}+ N\,,\qquad B \to h +\ell_{\alpha}+ N\,,
\end{equation}
where $h = K,\pi,\dots$ denotes a hadron. For the mass range $m_{N}\lesssim m_{K}-m_{l_{\alpha}}$ and the HNLs mixing with $\nu_{e,\mu}$, kaon decays provide an additional source,
\begin{equation}
    K^{\pm}\to \ell_{\alpha}^{\pm}+N\,.
\end{equation}
For the $\tau$ mixing case, the corresponding production mode is kinematically inaccessible.

The HNL decay modes relevant for displaced-decay searches include purely leptonic channels such as $N\to \ell_{\alpha}^{-}\ell_{\beta}^{+}\nu_{\beta}$, semileptonic charged channels such as $N\to \pi^{\pm}\ell_{\alpha}^{\mp}$, and neutral channels such as $N\to \pi^{0}\nu_{\alpha}$, together with the corresponding charge-conjugate modes~\cite{Bondarenko:2018ptm}.

Secondary kaons have been studied as a source of HNLs at SHiP in Ref.~\cite{Gorbunov:2020rjx}. In that work, the kaon population inside the target was obtained using a detailed \textsc{GEANT4} simulation \cite{GEANT4:2002zbu,Allison:2006ve,Allison:2016lfl}. The authors found that roughly $0.29$ ($0.07$) of $K^{+}$ ($K^{-}$) per proton-on-target decays in flight and may produce HNLs pointing to the SHiP decay volume, compared to produced yields of $8.01$ ($3.51$). The rest of the kaons are either absorbed through scatterings or decay at rest, producing the flux of isotropic HNLs. As the decay volume of SHiP covers a tiny fraction of the solid angle, the at-rest decays contribute negligibly.

The estimates of the HNL decay event rate were organized as follows. The authors simulated the decays of the in-flight kaons into HNLs, propagated the HNLs to the decay volume, and then simulated their decays. The trajectories of the two visible decay products were then required to intersect the cross-section of the end of the decay volume as a minimal decay product acceptance requirement, and each of them was required to have $p>1\,\gev$.

The study found that, despite the small angular acceptance of HNLs from secondary kaons, the large kaon flux and the reduced HNL boost may provide an additional contribution to the SHiP sensitivity compared with the contribution from charm and beauty meson decays alone.

Production from decays of secondary heavy mesons gives another contribution to the total HNL sensitivity~\cite{Ovchynnikov:2023cry}. However, as discussed above, the enhancement of heavy-meson production in the cascade is moderate, and the resulting LLP kinematics are much closer to the primary-production case. For this reason, the present work concentrates instead on lower-energy kaon sources.

Below, we consider the HNLs mixing with $\nu_{e}$ for numeric estimates. We comment on the mixing with $\nu_{\mu}$ in Sec.~\ref{sec:results}.

\subsection{Limitations of existing estimates}
\label{ssec:existing-studies-limitations}

The observable number of cascade-induced LLP decays is controlled by several distinct factors. In the long-lifetime regime, a useful schematic comparison between the primary and cascade production is
\begin{multline}
    \frac{N_{\rm ev,sec}}{N_{\rm ev,prim}}
    \sim
    \frac{N_{\rm X,prod,sec}}{N_{\rm X,prod,prim}}
    \times
    \frac{\epsilon_{\rm geom,dec,sec}}{\epsilon_{\rm geom,dec,prim}}
    \times
    \frac{\langle\gamma_{\rm X,sec}^{-1}\rangle}{\langle\gamma_{\rm X,prim}^{-1}\rangle}
    \\
    \times
    \frac{\epsilon_{\rm geom,det,sec}}{\epsilon_{\rm geom,det,prim}}
    \times
    \frac{\epsilon_{\rm reco,sec}}{\epsilon_{\rm reco,prim}}\,.
\end{multline}
Here, $N_{\rm X,prod}$ is the number of produced LLPs, $\epsilon_{\rm geom,dec}$ is the fraction of LLPs intersecting the decay volume, $\langle\gamma_X^{-1}\rangle$ accounts for the decay probability in the long-lifetime regime, $\epsilon_{\rm geom,det}$ is the probability that the visible decay products remain inside the relevant detector apertures, and $\epsilon_{\rm reco}$ is the probability that the event can be reconstructed with the required vertexing, pointing, invariant mass, and particle identification performance.

Existing cascade studies have addressed the first factors in this expression in considerable detail. In particular, they have estimated the secondary production rates, the angular and energy distributions of the produced LLPs, and the enhancement of the decay probability associated with the reduced boost. However, the last two factors, $\epsilon_{\rm geom,det}$ and $\epsilon_{\rm reco,sec}$, were treated with approximate prescriptions.

First, for the low-energy cascade LLPs, requiring the parent LLP to point toward the decay volume or a downstream plane does not by itself guarantee geometric detector acceptance for the decay products. The visible decay products must themselves remain inside the relevant detector apertures. For a two-body decay into light particles, the typical opening angle grows as the boost decreases, with $\Delta\theta\sim 2/\gamma_{X}$, with $\gamma_{X}$ being the LLP's $\gamma$ factor. As a result, the daughter-level acceptance can be strongly suppressed for cascade-produced LLPs. Qualitatively, the requirement for the decay products to be within the ECAL acceptance is 
\begin{equation}
\Delta\theta \lesssim \frac{\sqrt{A_{\rm ECAL}}}{\Delta z_{\rm X-ECAL}}\,,
\end{equation}
where $\Delta z_{\rm X-ECAL}>12\,\m$ is the distance between the coordinate of LLP decay to the entrance of the ECAL. For the LLPs decaying at the beginning (end) of the decay volume, this requirement leads to the criterion $\gamma_{X}>25$ ($\gamma_{X}>5$), eliminating the low-energy part of the LLP flux. 

Second, the reconstruction efficiency may introduce an additional suppression. Low-energy photons suffer from the worst reconstruction in the ECAL due to the degraded detector relative resolution, later showers, and worse achieved angular resolution. Low-momentum charged particles are more strongly deflected by the spectrometer magnet and may fail to leave enough usable tracking information for vertex and momentum reconstruction. This is particularly relevant for HNLs from secondary kaons, since many visible HNL decay modes contain charged particles. In exclusive channels such as $N\to\pi^\pm\ell^\mp$, the reconstruction also requires reliable momentum measurement and particle identification, or at least a controlled assignment of mass hypotheses.

The purpose of the present work is to quantify the impact of these effects for representative low-energy signals at SHiP. We do not address an additional complication associated with low-energy signals, namely, the possible change in the background composition and rejection power. We regard this as the next step in evaluating the SHiP detector performance for cascade-induced signatures, to be carried out once the signal is shown to survive daughter-level acceptance and detector-level reconstruction requirements.

\section{Our approach}
\label{sec:our-approach}

Our approach to calculating the event rates with the LLPs is the following. 

We use the fluxes of primary and secondary photons for ALPs and kaons for the HNLs as input to generate the angle-energy differential distribution of these LLPs. These are used to conduct two different cross-validated studies. Based on the semi-analytic approach from Ref.~\cite{Ovchynnikov:2023cry}, we estimate the impact of the geometric acceptance of the detector and the Monte Carlo truth selection criteria from Sec.~\ref{sec:ship-baseline-selection} on the number of events. Then, in the case of ALPs, where a more detailed investigation is needed (see Fig.~\ref{fig:Nevents-ratio}), we perform a detector-level simulation to accurately calculate reconstruction efficiency, focusing on the domain of low-energy events.

\subsection{Fluxes of primary and secondary SM particles}
\label{ssec:SM-fluxes}

\subsubsection{Photons}

We have generated the flux of the primary photons originating from the decays of $\pi^{0},\eta \to 2\gamma$ produced in the interactions of the proton beam with the target. For this purpose, we used calibrated \textsc{Pythia8} on the data from collisions of a 400 GeV proton beam with a thin target within 20\% precision, see Ref.~\cite{Dobrich:2019dxc} for details. The primary photons were used as an input in the \textsc{ALPETITE} tool from Ref.~\cite{Patrone:2025fwk} to generate the flux of secondary photons. The public \textsc{ALPETITE}/\textsc{PETITE} setup contains the molybdenum shower dictionaries needed for the validation run, whereas the current SHiP documents assume a tungsten target. We have therefore generated a dedicated tungsten \textsc{PETITE} dictionary. This is done by reusing the existing $400\,\,\mathrm{GeV}$ VEGAS adaptive phase-space maps and recomputing the integrated cross sections and rejection-sampling maxima for the tungsten material parameters. We keep photons with
\begin{equation}
E_{\gamma}>100\,\,\mathrm{MeV}\,,
\end{equation}
following the \textsc{ALPETITE} benchmark setup. The normalization of the cascade photons is fixed by the primary photon yield:
\begin{multline}
    w_{\rm prim}\approx 2\times \big(N_{\pi^{0}}\times {\rm Br}(\pi^{0}\to 2\gamma)\\ +N_{\eta}\times {\rm Br}(\eta\to 2\gamma)\big)\approx 8.8\,,
\end{multline}
where $N_{\pi^{0}},N_{\eta}$ are the amounts of mother mesons per proton-on-target:
\begin{equation}
    N_{\pi^{0}} \approx 4.2\,, \quad N_{\eta} \approx 0.5\,,
\end{equation}
and a factor of two comes from the decay mode of these mesons into a pair of photons.

\begin{figure*}[t!]
    \centering
    \includegraphics[width=0.5\linewidth]{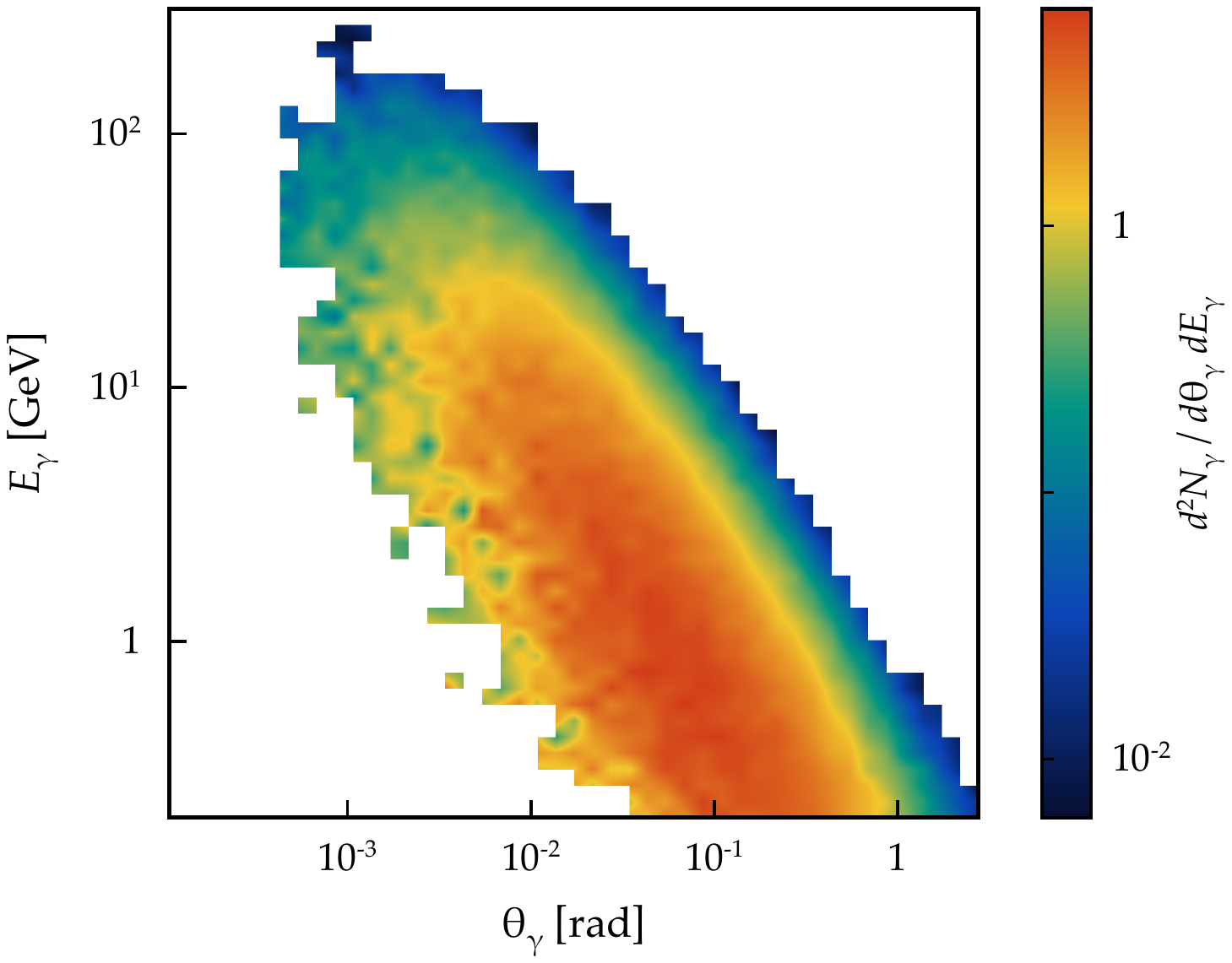}~\includegraphics[width=0.5\linewidth]{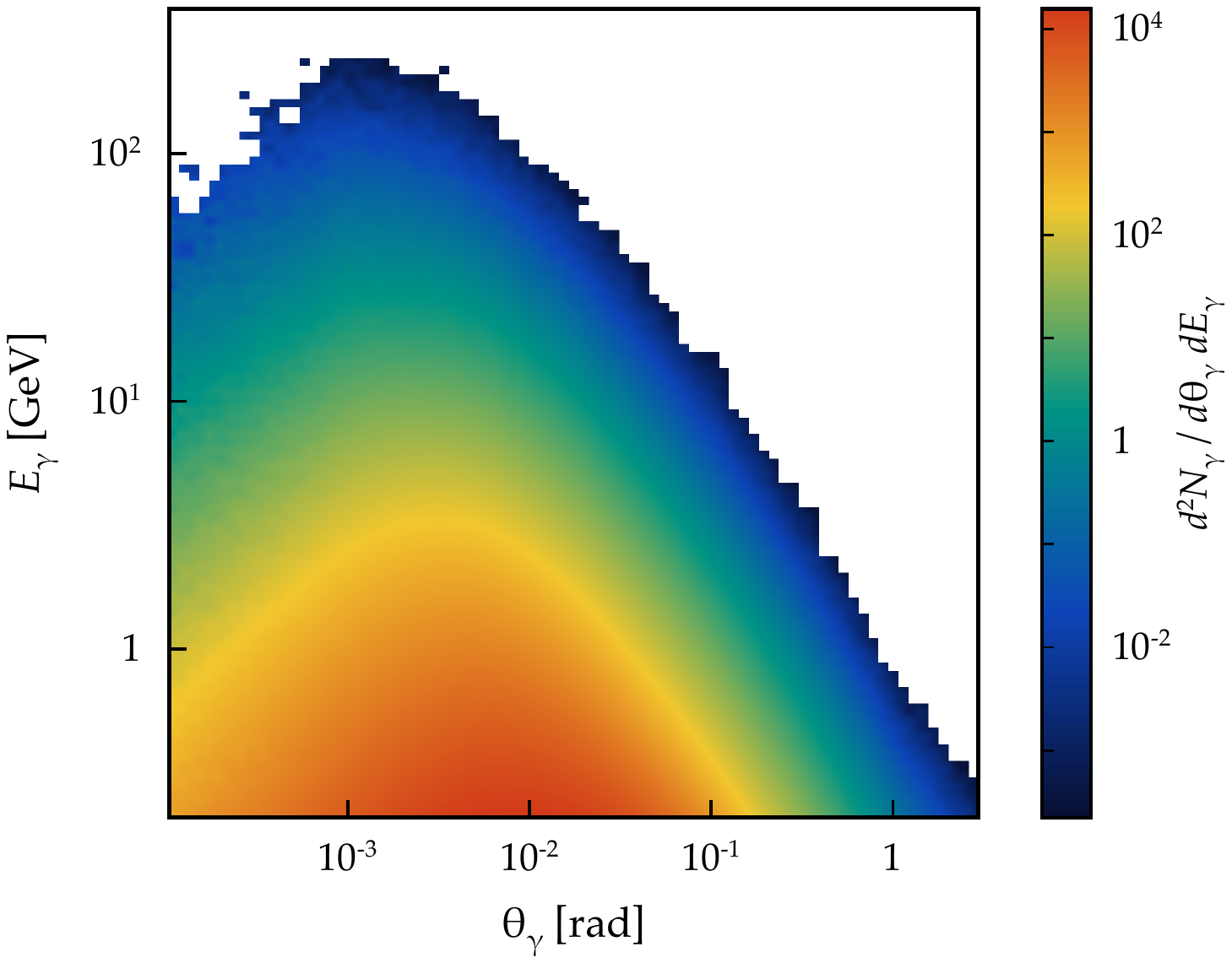}
    \caption{Angle-energy distribution $dN_{\gamma}/d\theta_{\gamma}dE_{\gamma}$ for primary (left) and secondary (right) photons at SHiP, normalized per proton-on-target. The SHiP target is assumed to be made of Tungsten. See text for details.}
    \label{fig:photon-flux}
\end{figure*}

The density plots of the angle-energy distribution of the primary and secondary photons are shown in Fig.~\ref{fig:photon-flux}. We see that the secondary photons are much softer on average; namely, the energy of $\gamma$s flying in the direction of the SHiP signal detectors (located at $\theta_{\gamma}<0.044\,\text{rad}$) is $\langle E_{\gamma,{\rm casc}}\rangle\approx 1\,\gev$ versus $\langle E_{\gamma,{\rm prim}}\rangle \approx 16.6\,\gev$. For the Primakoff conversion, $E_{\gamma}$ is an accurate proxy for the ALP energy, $E_{a}\approx E_{\gamma}$ (see relation~\eqref{eq:Egamma-Ea}), and hence this property translates to the primary and secondary ALPs.

\subsubsection{Kaons}

\begin{figure}[t!]
    \centering
    \includegraphics[width=\linewidth]{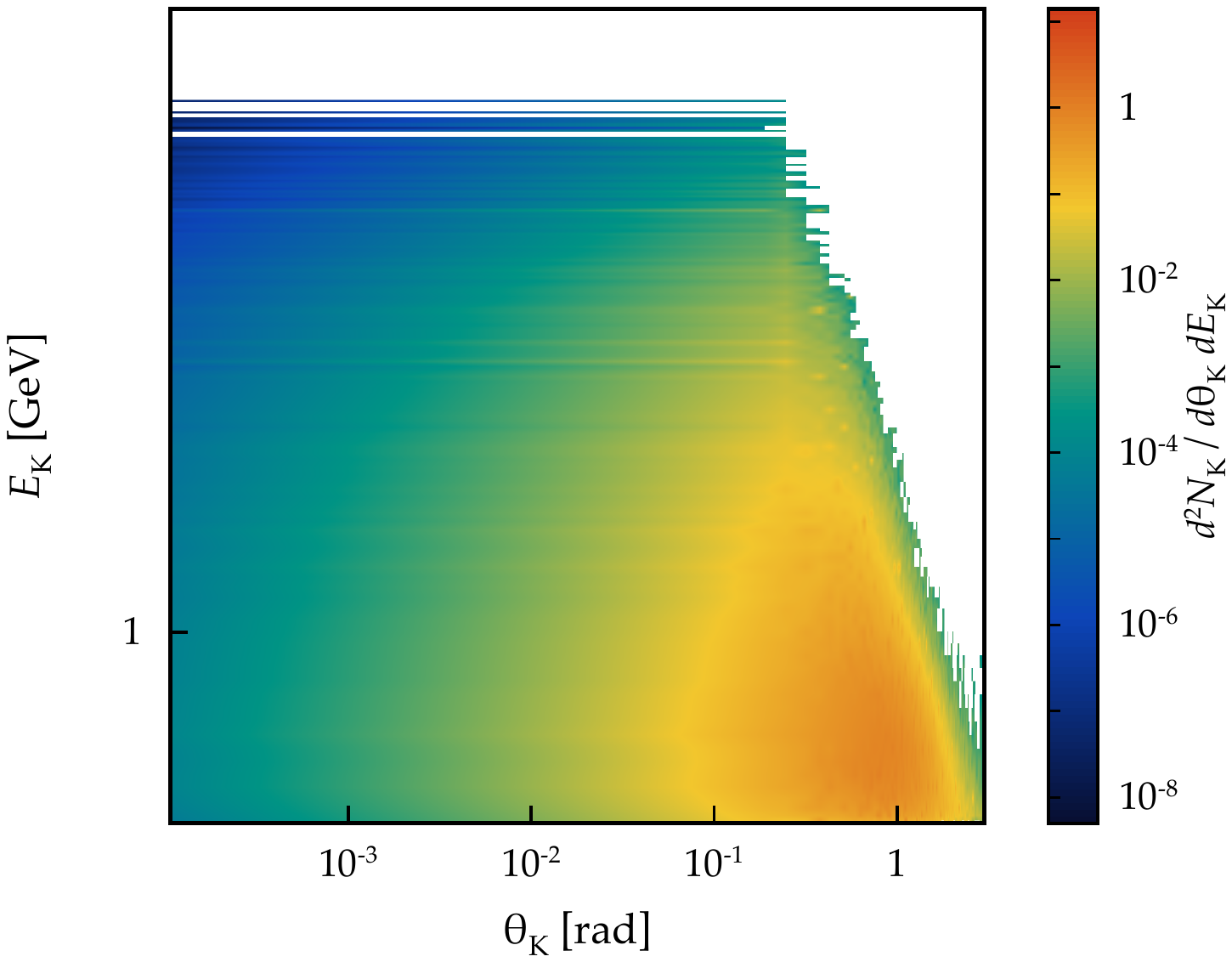}
    \caption{Angle-energy distribution $dN_{K}/d\theta_{K}dE_{K}$ for charged kaons $K = K^{+}+K^{-}$ at SHiP, normalized per proton-on-target, as obtained in Ref.~\cite{Gorbunov:2020rjx} assuming a hybrid Molybdenum-Tungsten SHiP target. The constant-like behavior of the energy distribution with the polar angle $\theta_{K}$ is caused by the extrapolation in the domain of small $\theta_{K}<\theta_{K,\rm min}\approx 0.1$, for which the binning of Ref.~\cite{Gorbunov:2020rjx} was not fine enough. Namely, $dN_{K}/d\theta_{K}dE_{K}\sim \frac{\theta_{K}}{\theta_{\rm K,min}}\left(\frac{dN_{K}}{d\theta_{K}dE_{K}}\right)\bigg|_{\theta_{K}\theta_{\rm K,min}}$. See text for details.}
    \label{fig:kaon-flux}
\end{figure}

For the decaying kaon flux and multiplicities, we will use the results of Ref.~\cite{Gorbunov:2020rjx}. The authors prepared the flux of charged kaons assuming the Molybdenum-Tungsten target. The flux may be somewhat smaller for the current SHiP target design, which is made solely of Tungsten. Nevertheless, this is enough for obtaining qualitative conclusions about the contribution of the cascade HNLs, as we will show in Figs.~\ref{fig:energy-distribution-HNL},~\ref{fig:Nevents-ratio}.

The primary kaons have a small probability of decaying before absorption compared to the secondary ones (due to the high energies of the primaries). For this reason, we effectively treat all the kaons decaying in-flight (and hence HNLs from them) as cascade particles.

The angle-energy distribution of all the kaons decaying in-flight is shown in Fig.~\ref{fig:kaon-flux}. Compared to the photons, the kaons are much more isotropic and softer. This is caused by a strong kaon-energy dependence of the decay probability. Therefore, most of the HNLs would be produced close to isotropically.

\subsection{Generating the LLP flux}
\label{sec:geom-acceptance}

Having the events with the primary and secondary photons, as well as kaons, we tabulate them and use them as an input to the \textsc{SensCalc} tool~\cite{Ovchynnikov:2023cry}.\footnote{Available on Zenodo~\cite{SensCalc} and \faGithub~\cite{SensCalc-GitHub}.} \textsc{SensCalc} is based on a semi-analytic method to compute the number of events with different LLPs at various experiments, taking into account the geometry of the decay volume, detector, and the selection criteria of the decay events based on the Monte-Carlo-truth level cuts.

The differential rate of the produced ALPs is
\begin{multline}
    \frac{dN_{\rm a,prod}}{d\theta_{\gamma}\,dE_{\gamma}\,dq^{2}}
    \simeq
    \frac{1}
         {\sigma_{\gamma,{\rm abs}}}\frac{dN_{\gamma}}{d\theta_{\gamma}\,dE_{\gamma}}
    \\
    \times
    \left[
    \frac{d\sigma_{\gamma Z\to aZ}}{dq^{2}}
    +
    Z\,\frac{d\sigma_{\gamma p\to ap}}{dq^{2}}
    \right]\,.
    \label{eq:Nevents-ALP}
\end{multline}
Here, the quantity $\sigma_{\gamma,{\rm abs}}$ denotes the effective photon absorption cross section
in the material, $q^{2} = -(p_{a}-p_{\gamma})^{2}$ is the momentum transfer to the scattering target. Finally, the two terms in brackets of Eq.~\eqref{eq:Nevents-ALP} describe coherent scattering on the nucleus and incoherent scattering on the individual protons. For the differential Primakoff cross sections, we use the expressions of Ref.~\cite{Dobrich:2019dxc},
including the atomic screening factor for the nuclear target.

The distribution above must be connected to the ALP polar angle $\theta_{a}$ and energy $E_{a}$ distributions. The previous implementation in \textsc{SensCalc} (Ref.~\cite{Ovchynnikov:2023cry}) used the collinear approximation $E_{a}\approx E_{\gamma}, \theta_{a}\approx \theta_{\gamma}$. It is a reliable description for photons with energy $E_{\gamma}\gg m_{a}$ but may break down for the soft cascade photons. This motivated an alternative strategy to obtain the ALP distribution: mapping the variables event by event for scattering processes on a target $T=Z,p$ of mass $m_T$.

Namely, for the given $q^{2}$, we determine the ALP energy is
\begin{equation}
    E_a = E_\gamma - \frac{q^2}{2m_T}\,.
    \label{eq:Egamma-Ea}
\end{equation}
In practice, the second term may often be neglected even for the light ALPs. The ALP angle $\psi_T$ with respect to the incoming photon direction is given by
\begin{equation}
    \cos(\psi_T) =
    \frac{2E_\gamma E_a - m_a^2 - q^2}
         {2E_\gamma\sqrt{E_a^2-m_a^2}}\,,
\end{equation}
with $|\cos(\psi_{T})|<1$. The azimuthal angle $\phi$ around the photon direction is taken to be uniformly distributed in $[0,2\pi)$. Therefore, if the incoming photon has polar angle $\theta_\gamma$ with respect to the beam axis, the ALP polar angle is
\begin{equation}
    \cos\theta_a =
    \cos\theta_\gamma\cos\psi_T
    +
    \sin\theta_\gamma\sin\psi_T\cos\phi .
\end{equation}
Using this transformation, the weighted distribution in $(\theta_\gamma,E_\gamma,q^2)$ from Eq.~\eqref{eq:Nevents-ALP} is converted into the final double-differential ALP distribution:
\begin{equation}
\frac{dN_{\rm a,prod}}{d\theta_a\,dE_a} \equiv N_{\rm a,prod}\times f_{a}(\theta_{a},E_{a})\,,
\end{equation}
where $N_{\rm a,prod}$ is the total number of produced ALPs, and $f_{a}(\theta_{a},E_{a})$ the ALP angle-energy distribution normalized to unity.

For the HNLs, we straightforwardly simulate their production in kaon decays via the processes $K\to N+l_{\alpha}$, where $\alpha$ corresponds to the mixing angle $U_{\alpha}$~\cite{Bondarenko:2018ptm}.

\begin{figure*}[t!]
    \centering
    \includegraphics[width=0.5\linewidth]{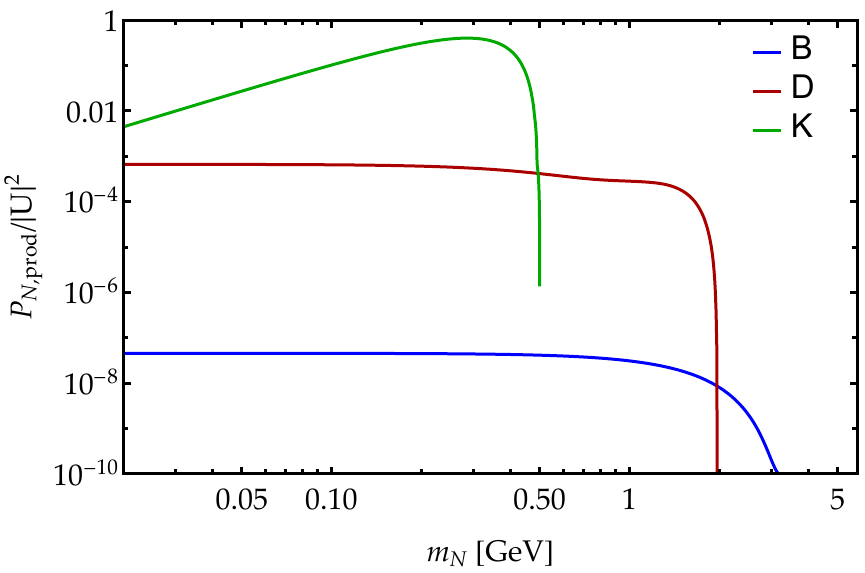}~\includegraphics[width=0.5\linewidth]{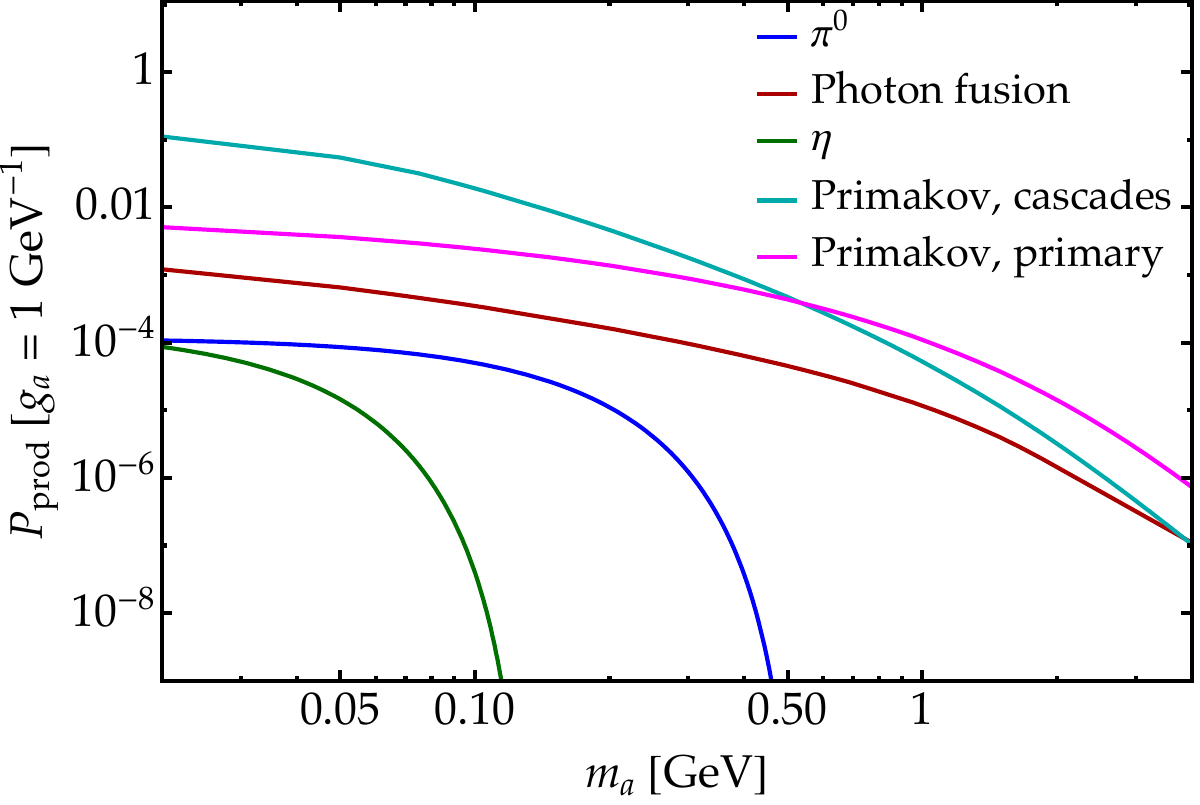}
    \caption{Overall production probability (per proton-on-target) of HNLs mixing with $\nu_{e}$ (left panel) and photophilic ALPs (right panel). For the ALPs, we have considered the secondary photons with energies $E_{\gamma}>100\,\mev$. For the cascade HNLs, the shape of the mass dependence of the production probability is mainly caused by the mass dependence of the branching ratio of the process $K\to N + e$ (chirality suppression at small $m_{N}\ll m_{K}$), while for the ALPs it is driven by the steep rise of the photon flux toward lower $E_{\gamma}$ (see Fig.~\ref{fig:photon-flux}), and approximate energy correspondence $E_{a}\approx E_{\gamma}$ in the Primakoff process (Eq.~\eqref{eq:Egamma-Ea}).}
    \label{fig:overall-production-fluxes}
\end{figure*}

To demonstrate the potential importance of the cascade production, we show the integrated yields of the ALPs and HNLs per proton-on-target in Fig.~\ref{fig:overall-production-fluxes}. It is clearly seen that before the geometric acceptance and event reconstruction are taken into account, the secondary production channels may dominate the flux of LLPs $X$ with mass $m_{X}\lesssim 1\,\gev$. In particular, recalling the discussion in Sec.~\ref{ssec:existing-studies-limitations}, even if geometric acceptance of the decay volume somehow suppresses the event rate with cascade particles (as it may be the case for the HNLs from kaons, recall Fig.~\ref{fig:kaon-flux}), this is partially compensated by the $1/\gamma_{X}$ scaling of the decay probability.

\subsection{Semi-analytic event rate calculation}
\label{ssec:semi-analytic-calculation}
Once the cascade ALP and HNL fluxes are tabulated, we compute the event rate with their decays using \textsc{SensCalc}. Namely, for an LLP $X$ with mass $m_{X}$ and coupling to SM particles $g_{X}$, the number of decay events is~\cite{Ovchynnikov:2023cry}
\begin{multline}
    N_{\rm ev}(m_{X},g_{X}) = N_{\rm X,prod}(m_{X},g_{X})\\ \times \int d\theta_{X}dE_{X}dz_{X}\, f_{X}(\theta_{X},E_{X})\frac{\Delta \phi(\theta_{X},z_{X})}{2\pi} \\ \times
\frac{dP_{\rm decay}}{dz_{X}}\epsilon_{\rm dec}(\theta_{X},E_{X},z_{X})
\,.
\label{eq:Nevents-semi-analytic}
\end{multline}
Here, $N_{\rm X,prod}$ is the total number of produced LLPs. $\theta_{X},E_{X},z_{X}$ are the polar angle, energy, and the longitudinal coordinate of the decaying $X$ with respect to the target. $f_{X}(\theta_{X},E_{X})$ is the angle-energy distribution of the produced LLPs, normalized to unity. $\Delta \phi(\theta_{X},z_{X})$ is the fraction of the azimuthal angle covered by the decay volume. $dP_{\rm decay}/dz_{X}$ is the differential decay probability of the LLP:
\begin{equation}
  \frac{dP_{\rm decay}}{dz_{X}} =\frac{\exp\left[-\frac{z_{X}}{c\tau_{X}\sqrt{\gamma_{X}^{2}-1}\cos(\theta_{X})}\right]}{c\tau_{X}\sqrt{\gamma_{X}^{2}-1}\cos(\theta_{X})}\,,
  \label{eq:dPdecaydz}
\end{equation}
with $\tau_{X}(m_{X},g_{X})$ being the $X$'s lifetime. Finally, $\epsilon_{\rm dec}$ is the decay acceptance -- the fraction of the events with a decaying LLP where at least two decay products are within the geometric acceptance of the detector and also satisfy other cuts, such as energy, $p_{T}$ and transverse spatial separation.

Let us discuss the calculation of $\epsilon_{\rm dec}$ in more detail. It is obtained for each point $(\theta_{X},E_{X},z_{X},\{\phi_{X}\})$ by simulating decays of $X$ into all allowed final states:
\begin{multline}
    \epsilon_{\rm dec}(\theta_{X},E_{X},z_{X}) \\ \equiv \sum_{f}\sum_{\phi_{X,i}\in \{\phi_{X}\}}\frac{\epsilon^{(f)}_{\rm dec}(\theta_{X},E_{X},z_{X},\phi_{X,i})}{N_{\phi}} \times \rm{Br}(X\to f)\,.
\end{multline}
Here, $\{\phi_{X}\}$ denotes the set of random $N_{\phi}$ azimuthal angles for which the decay point $\mathbf{r}_{X}(\theta_{X},z_{X},\phi_{X})$ is inside the decay volume. The individual acceptance $\epsilon^{(f)}_{\rm dec}(\theta_{X},E_{X},z_{X},\phi_{X,i})$ is obtained by simulating a large number of decays of $X$ (1000 by default), and may take a value \begin{equation}
0\leq\epsilon^{(f)}_{\rm dec}(\theta_{X},E_{X},z_{X},\phi_{X,i})\leq 1\,.
\end{equation}
Finally, $f$ is the concrete final state, with $\rm{Br}(X\to f)$ the corresponding branching ratio.

The trajectories of decay products are assumed to be unperturbed by medium interaction throughout the propagation, except for deflection caused by the spectrometer's magnetic field, which is implemented by a kick obtained through the Lorentz force at the end of the magnetized region.  

Additionally, the approach of Eq.~\eqref{eq:Nevents-semi-analytic} may accommodate the reconstruction efficiency map $\epsilon_{\rm rec}(\theta_{X},E_{X},z_{X})$, which is obtained by fitting the realistic detector response simulations, similarly to how this is done in \textsc{EvtGen} \cite{Lange:2001uf}.

For the calculation of the production and decay rates of HNLs and ALPs, we follow Refs.~\cite{Gorbunov:2007ak,Bondarenko:2018ptm,Ovchynnikov:2023cry}.

Within its purpose, the semi-analytic approach is very accurate and has been tested against various methods, including the \textsc{ALPINIST} framework for ALPs~\cite{Jerhot:2022chi} and the full SHiP simulation framework for dark photons and HNLs (see details in Ref.~\cite{Ovchynnikov:2023cry}).

We consider three types of decay event selection $\epsilon_{\rm dec}$ as summarized in Table~\ref{tab:semi-analytic-event-classification}.

\begin{table*}[t]
\centering
\small
\renewcommand{\arraystretch}{1.0}
\setlength{\tabcolsep}{4pt}

\newcommand{\tcell}[2]{%
  \begin{minipage}[t]{#1}%
  \raggedright
  #2%
  \end{minipage}%
}

\caption{Event selections used in the semi-analytic estimate by Eq.~\eqref{eq:Nevents-semi-analytic} (for the detector-level simulation, see Sec.~\ref{sec:signal-reconstruction} instead). The LLP is always required to decay inside the SHiP fiducial decay volume; the three cases differ by the requirements imposed on its visible decay products.}
\label{tab:semi-analytic-event-classification}

\begin{tabular}{llll}
\toprule
\tcell{0.15\textwidth}{\textbf{Selection}} &
\tcell{0.21\textwidth}{\textbf{Decay-volume requirement}} &
\tcell{0.27\textwidth}{\textbf{Daughter-level requirement}} &
\tcell{0.27\textwidth}{\textbf{Meaning for $a\to\gamma\gamma$}} \\
\midrule

\tcell{0.15\textwidth}{$\epsilon_{\rm dec}\equiv 1$} &
\tcell{0.21\textwidth}{The LLP decay vertex lies inside the SHiP fiducial decay volume.} &
\tcell{0.27\textwidth}{No requirement is imposed on the decay products.} &
\tcell{0.27\textwidth}{All LLP decays inside the fiducial volume are counted, independently of the photon energies, directions, or separation.} \\
\addlinespace[1.1em]

\tcell{0.15\textwidth}{Geom only} &
\tcell{0.21\textwidth}{The LLP decay vertex lies inside the SHiP fiducial decay volume.} &
\tcell{0.27\textwidth}{At least two visible decay products must reach the downstream detector aperture.} &
\tcell{0.27\textwidth}{Both photons are required to lie within the aperture of the end of the ECAL.} \\
\addlinespace[1.1em]

\tcell{0.15\textwidth}{Baseline} &
\tcell{0.21\textwidth}{The LLP decay vertex lies inside the SHiP fiducial decay volume.} &
\tcell{0.27\textwidth}{The baseline SHiP selection is imposed at MC-truth level, Eqs.~\eqref{eq:selection-ship-baseline-1}--\eqref{eq:selection-ship-baseline-4}.} &
\tcell{0.27\textwidth}{Requirements on the photon energies and their transverse separation at the ECAL.} \\

\bottomrule
\end{tabular}
\end{table*}

\subsection{Detector-level study}
\label{sec:signal-reconstruction}

In this subsection, we focus on the ALP case. To study the impact of detector-level effects on signal reconstruction, we export the tabulated ALP angle-energy distribution and production probabilities to \textsc{EventCalc-SHiP}~\footnote{Available on \faGithub~\cite{EventCalc}.}. This tool samples the full decay-event kinematics of LLPs decaying inside the SHiP decay volume. Its inputs consist of the tabulated angle-energy distribution of the produced LLPs, together with the model-dependent decay information: the LLP lifetime as a function of mass and coupling, the decay branching ratios, and, for three-body decays, the squared matrix elements governing the decay-product distributions in the LLP rest frame.

Using tabulated LLP production distributions is useful because it separates the production calculation from the decay and detector-simulation steps. In this way, one does not need to implement every production channel directly inside the event generator. This is especially important for cascade-induced signals, where the production calculation may involve dedicated shower simulations, material-dependent cross sections, and model-specific reweighting procedures. The tabulated interface, therefore, keeps the event-generation code modular and easier to maintain.

Predictions of \textsc{EventCalc} agree with \textsc{SensCalc} at the level of 5\%, both in terms of the integrated quantities and the differential event distributions. However, given the sophisticated method of calculating the ALP flux, we additionally tested it against \textsc{ALPETITE}, namely comparing the output of \textsc{EventCalc} with Fig. 3 of Ref.~\cite{Patrone:2025fwk}, finding an excellent agreement (see Appendix~\ref{app:cross-check-ALPETITE}). 

The output -- the information about the decay vertex coordinates, kinematics, and types of the decay products -- may be passed to a detector simulation. Our focus here is on the ECAL performance, and below, we describe in detail the standalone \textsc{GEANT4} simulation to study it.

\subsubsection{GEANT4 simulation framework}
\label{ssec:simulation}

The tabulated \textsc{EventCalc-SHiP} data is propagated into a custom simulation faithfully representing the SHiP ECAL and HCAL (although only the former is used in this work).

The detector geometry is described using \textsc{GeoModel} \cite{Bandieramonte:2021ast} and passed onto \textsc{GEANT4}, which then handles propagation of particles in matter using the \textsc{FTFP\_BERT} physics list. The raw \textsc{GEANT4} hits are digitized to most closely mimic detector response, converting energy deposition into parametrized photon distributions accounting for scintillation light, photon trapping, photon escape, and photodetection by the SiPMs. The obtained signals are then converted into ADC counts, assuming usage of the KLauS ASIC as a front end based on previous experience operating prototypes of the detector \cite{Climescu:2025kdj,Briggl:2016tlo}.

The digitized detector response is then converted into reconstructed hits, clustered, and the information is used to reconstruct the event using an early version of the calorimeter reconstruction procedures, which will be documented in a future publication. Two clusters are expected per event when two photons are within detector acceptance.

\subsubsection{Event vertex and invariant mass reconstruction}
\label{ssec:reconstruction}

The cluster energy is computed and calibrated in units of minimum-ionizing-particle depositions (MIPs) using a muon particle gun sample as a reference. The spatial distribution of hits in each cluster is then analyzed using a principal component analysis (PCA) \cite{Wind:1984cx}. The shower direction is estimated from the eigenvector of the hit-position covariance matrix corresponding to the largest eigenvalue. The cluster directions and positions are then used to compute a point of closest approach (POCA) vertex. The energies and directions are also used to compute the two-photon invariant mass:

\begin{equation}
    m_{\gamma\gamma}^{2}= 2 E_0 E_1 (1 - \cos\ \theta_\text{open})\,,
\end{equation}

with $\theta_\text{open}$ the opening angle between the two clusters with energies $E_{0}$, $E_{1}$. The photon cluster information is also used to derive an impact parameter of the vertex to the target, which may be used to reject background. Said background is nonetheless expected to be negligible and primarily emerging from deep-inelastic-scattering processes within the decay vessel walls. As a result, events are deemed to pass signal level cuts with 3 levels of selection being defined:
\begin{itemize}
    \item The weak selection requires two reconstructed clusters found by the detector. This selection does not assume any vertex reconstruction. It effectively corresponds to the pure geometric requirement for the two photons to be within the ECAL and be separated enough to form non-overlapping electromagnetic showers.
    \item The medium selection extends the weak selection by requiring the reconstructed vertex to lie inside a reduced fiducial decay volume. The reduced volume is obtained by shrinking the nominal decay volume by $1\sigma$ vertex resolution in each coordinate. As an example, if the vertex resolution for a given ALP scenario is $\sigma_{\text{vtx}} = \sigma_{\text{vty}} = \SI{5}{\centi\meter}$ and $\sigma_{\text{vtz}}=\SI{5}{\meter}$ in $z$, the fiducial volume will be defined as the decay volume's pyramidal frustum extending from the $0.9\times\SI{2.9}{\meter\squared}$ plane centered around 0 in $z\sim \SI{33}{\meter}$ to $3.9\times\SI{5.9}{\meter\squared}$ in z$\sim\SI{83}{\meter}$ from $z\sim\SI{38}{\meter}$ to $z\sim\SI{78}{\meter}$. No distance of closest approach cut was implemented as it was found to have little impact after the vertex selection.
    \item A tight selection which further expands on the medium one, requiring in addition that the reconstructed impact parameter to the target be smaller than \SI{1}{\meter} to account for the relatively lower resolution in electromagnetic shower vertexing. This selection is only expected to be necessary in the real experiment if the background is found to be significant and is thus deemed very conservative.
\end{itemize}

The impact-parameter threshold of the tight selection is more stringent than the baseline criterion~\eqref{eq:selection-ship-baseline-3}, ${\rm IP}<2.5\,\m$, used in the semi-analytic study of Sec.~\ref{ssec:semi-analytic-calculation}. We adopt the tighter value throughout the detector-level analysis as a conservative choice, so that the impact-parameter penalties reported in Sec.~\ref{ssec:selection-cuts} should be understood as upper estimates.

\section{Results and discussion}
\label{sec:results}

\subsection{Diagnostic effects of low-energy cascade kinematics}
\label{ssec:diagnostics}

\begin{figure}[t!]
    \centering
    \includegraphics[width=\linewidth]{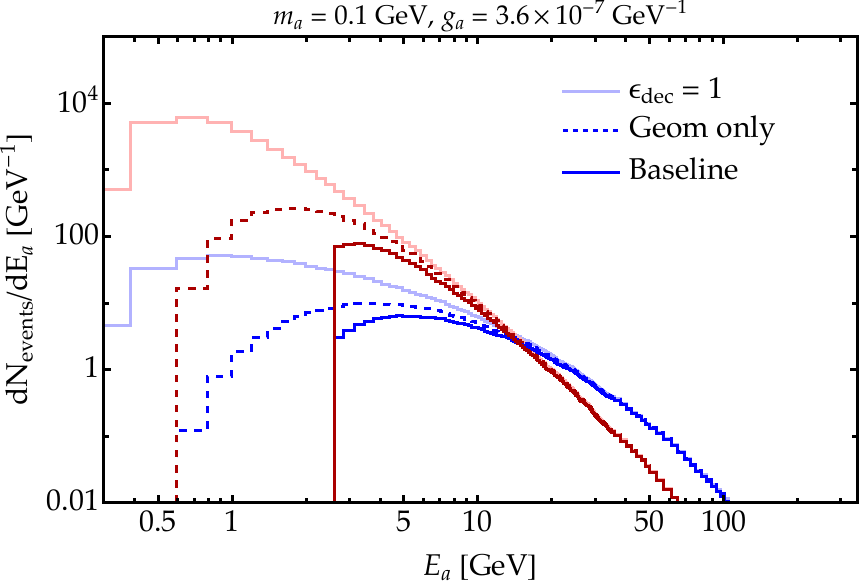} \\
    \includegraphics[width=\linewidth]{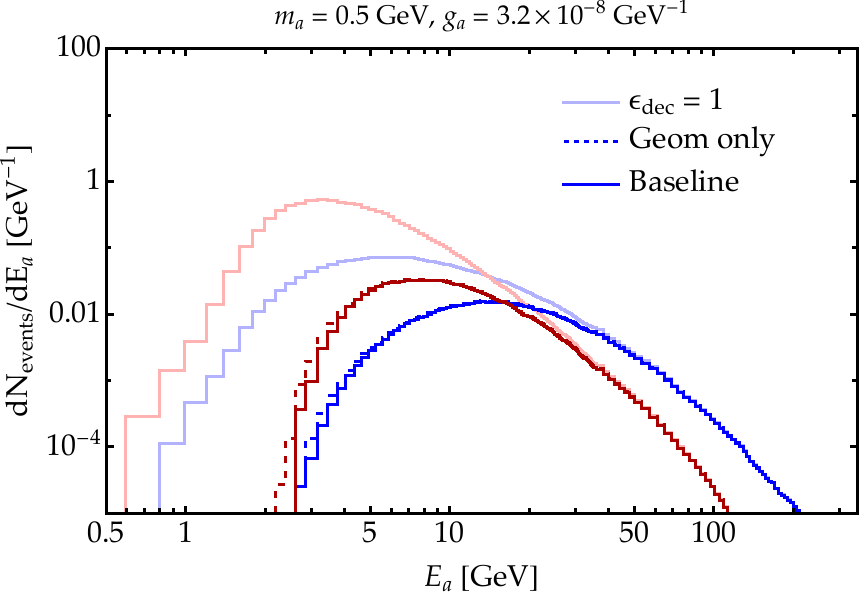}
    \caption{The distribution of the ALP decay events in the ALP energy. Top panel: $m_{a} = 100\,\mev$; bottom panel: $m_{a}=500\,\mev$. The ALP coupling $g_{a}$ is fixed in a way such that $c\tau_{a} = 300\,\m$. The blue lines show the contribution from the primary ALP fluxes, including the Primakoff process, photon fusion, and decays of $\pi^{0},\eta$ mesons, whereas the red lines show the contribution from the cascade ALP flux originating from the Primakoff process. For each color, pale, dashed, and dark stylings show, correspondingly, the number of events calculated assuming ``$\epsilon_{\rm dec} = 1$'', ``Geom only'', and ``Baseline'' selection criteria summarized in Tab.~\ref{tab:semi-analytic-event-classification}.}
    \label{fig:energy-distributions-ALP}
\end{figure}

Cascade production may strongly increase the number of produced LLPs, but the resulting LLPs are typically much softer than those from primary production. This has two immediate consequences for displaced-decay searches. First, the visible decay products have larger opening angles, which may prevent them from simultaneously entering the downstream detector acceptance. Second, their lower energies make the reconstruction of individual objects and vertices more challenging. In this subsection, we isolate these effects before combining them into the total event yield.

\begin{figure}[t!]
    \centering
    \includegraphics[width=\linewidth]{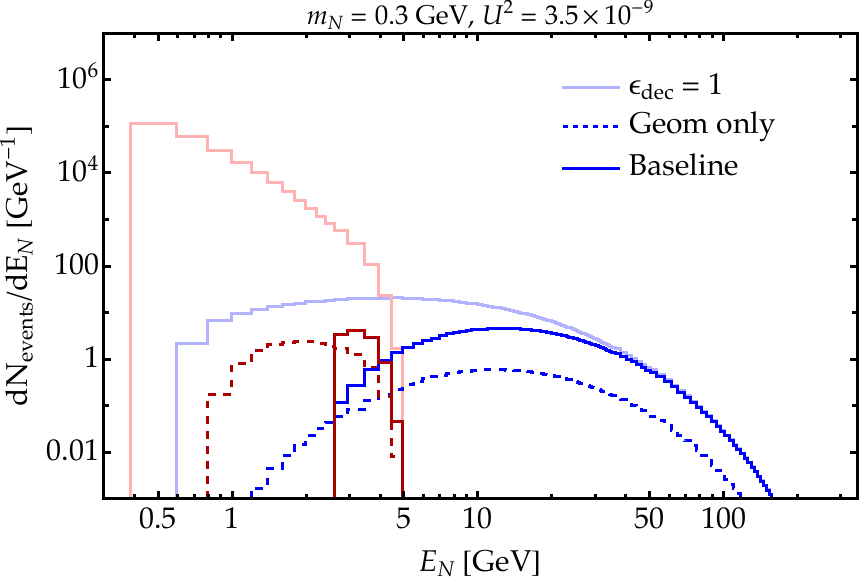}
    \caption{The distribution of the HNL decay events in the HNL energy for $m_{N}=300\,\mev$, assuming the mixing with electron neutrinos, for the lifetime $c\tau_{N}\approx 2.3\times 10^{7}\text{ m}$. The meaning of the curves is the same as in Fig.~\ref{fig:energy-distributions-ALP}, modulo the different sources of primary and cascade HNLs, as specified in the text.}
    \label{fig:energy-distribution-HNL}
\end{figure}

\subsubsection{Semi-analytic estimates}
\label{sec:discussion-semi-analytic}

We start with the semi-analytic analysis by Eq.~\eqref{eq:Nevents-semi-analytic}, and apply it to the ``$\epsilon_{\rm dec} = 1$'', ``Geom only'', and ``Baseline'' selections on decay products summarized in Tab.~\ref{tab:semi-analytic-event-classification}.

\begin{figure*}[!t]
\centering
\includegraphics[width=0.32\linewidth]{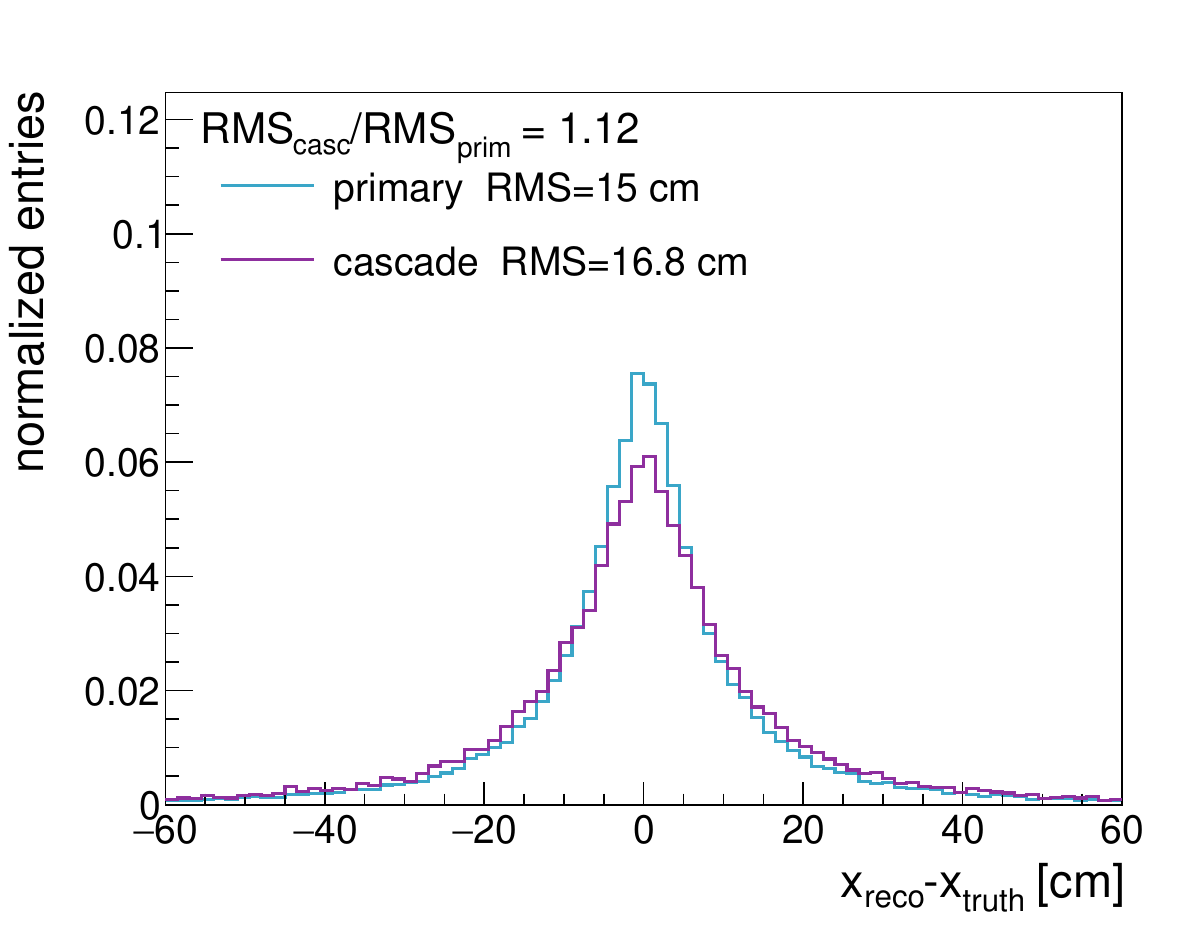}
\includegraphics[width=0.32\linewidth]{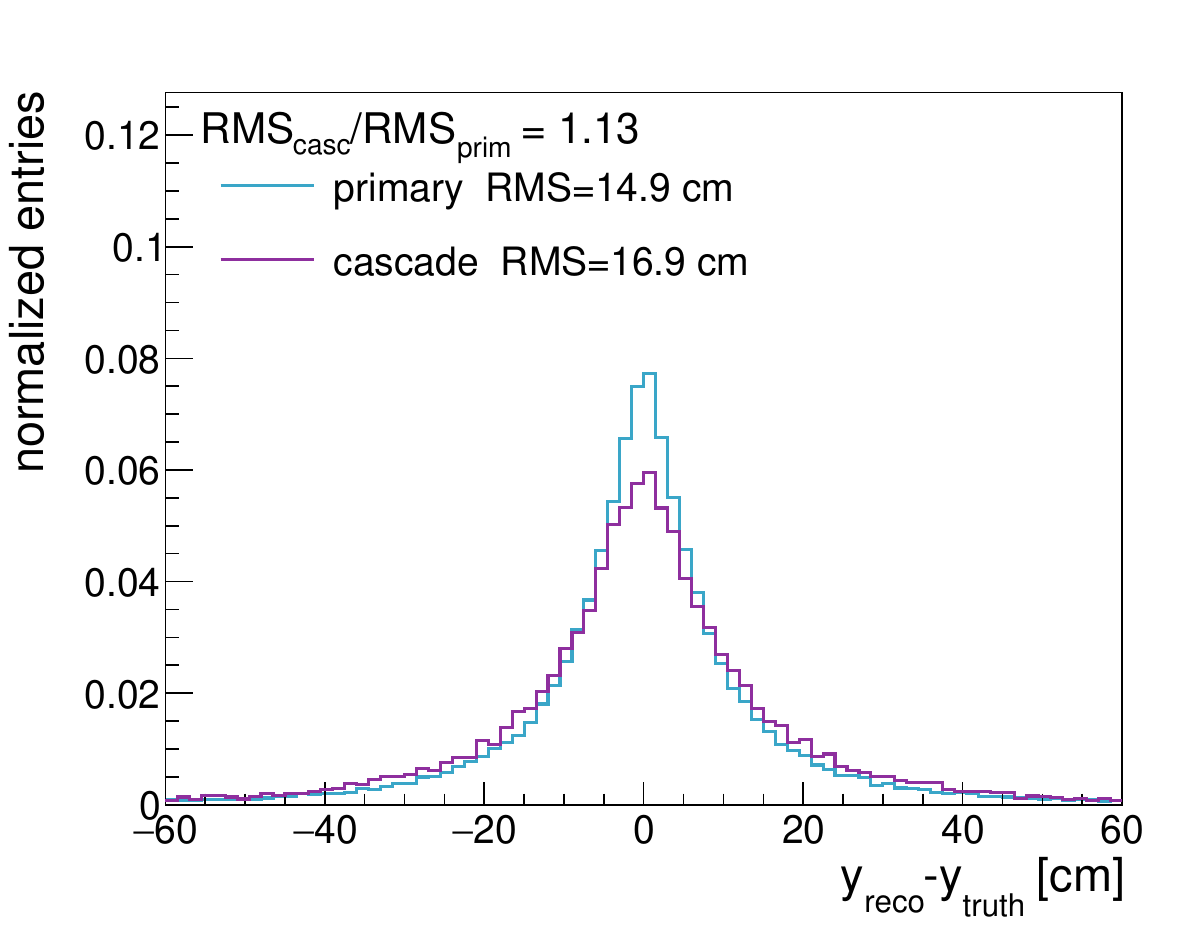}
\includegraphics[width=0.32\linewidth]{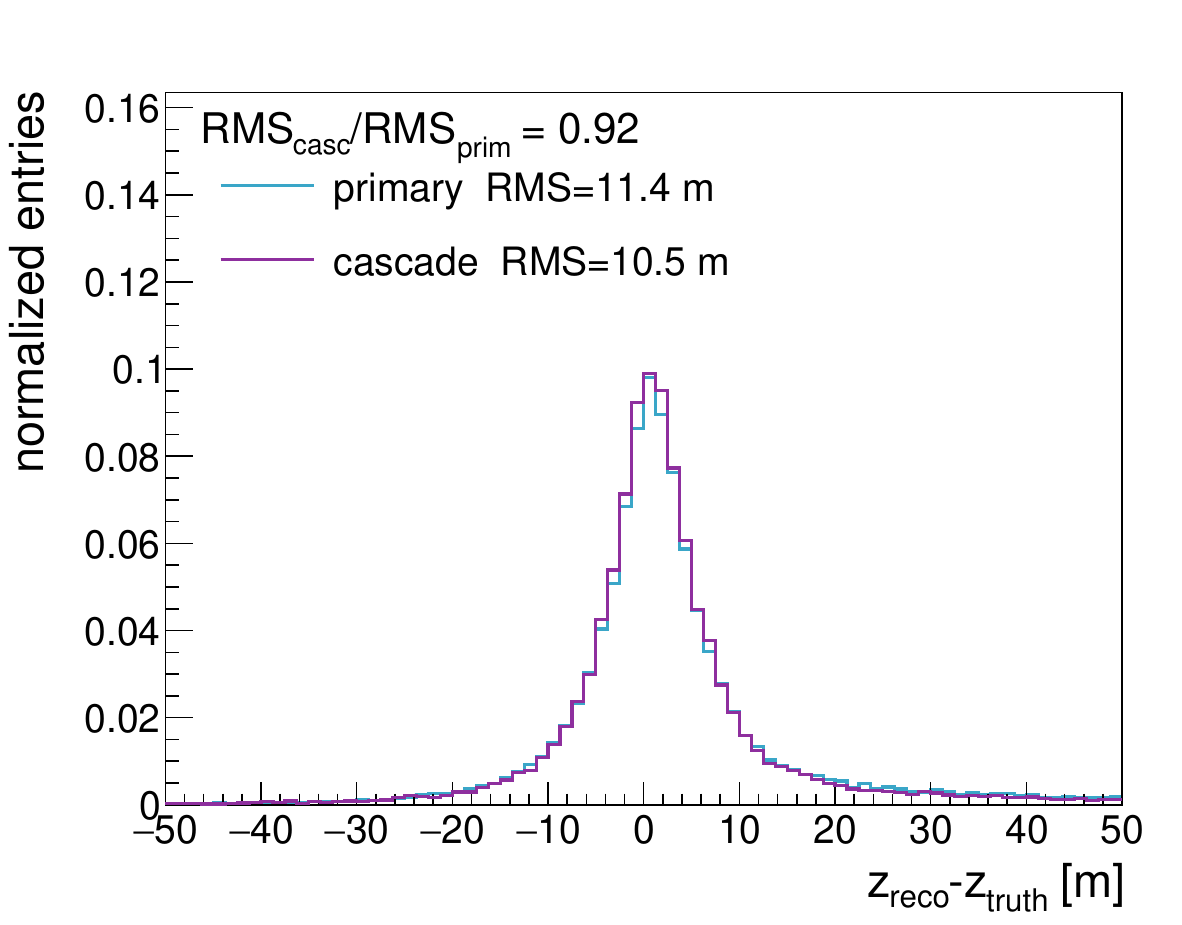}

\includegraphics[width=0.32\linewidth]{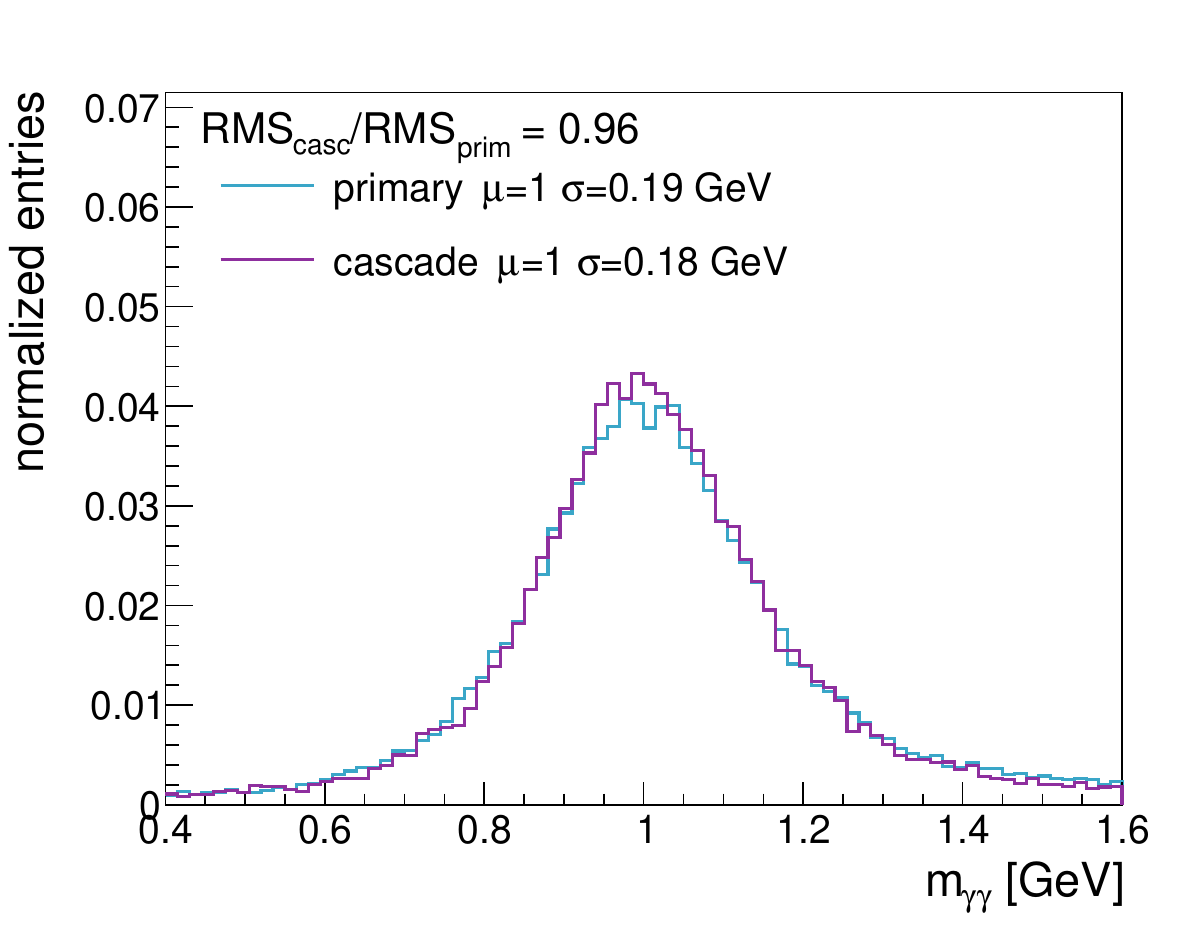}
\includegraphics[width=0.32\linewidth]{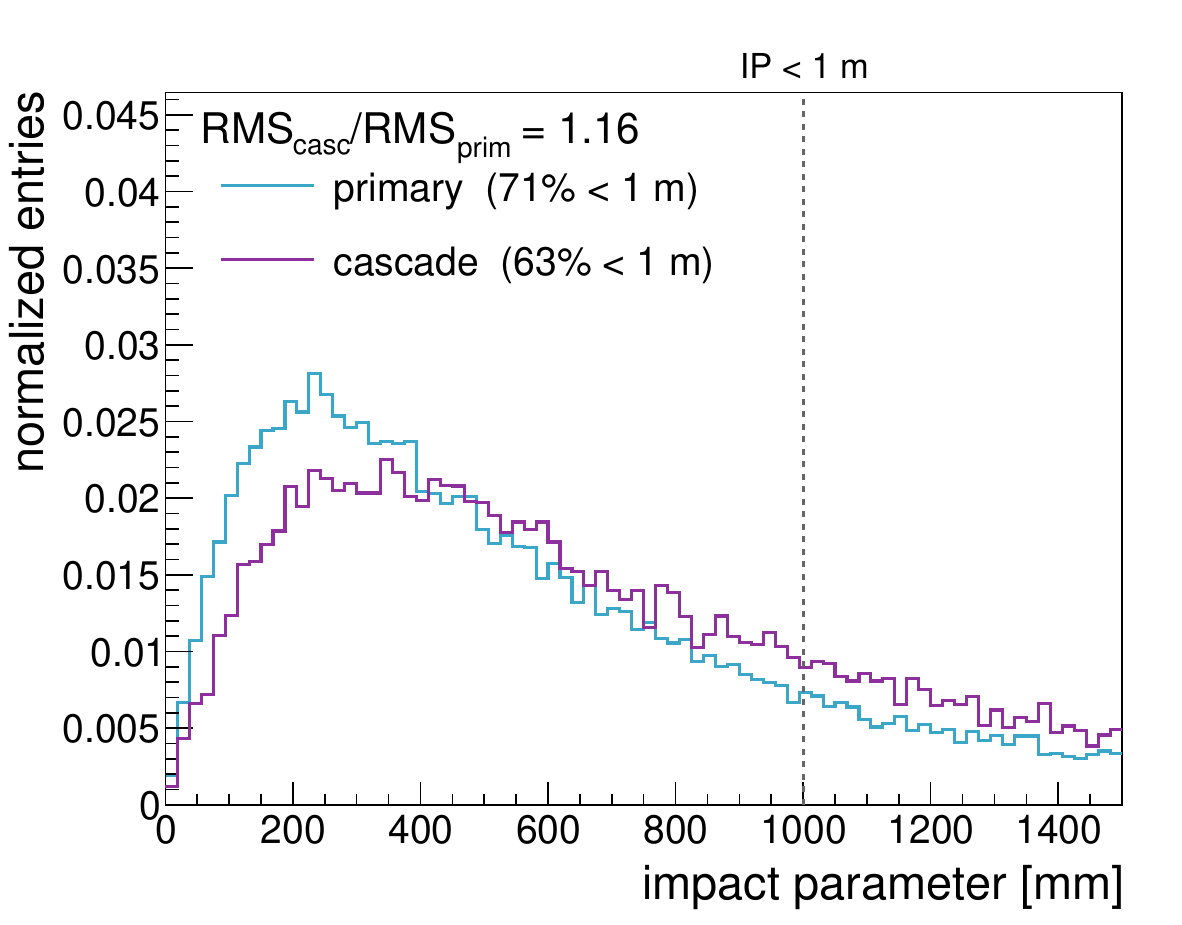}
\includegraphics[width=0.32\linewidth]{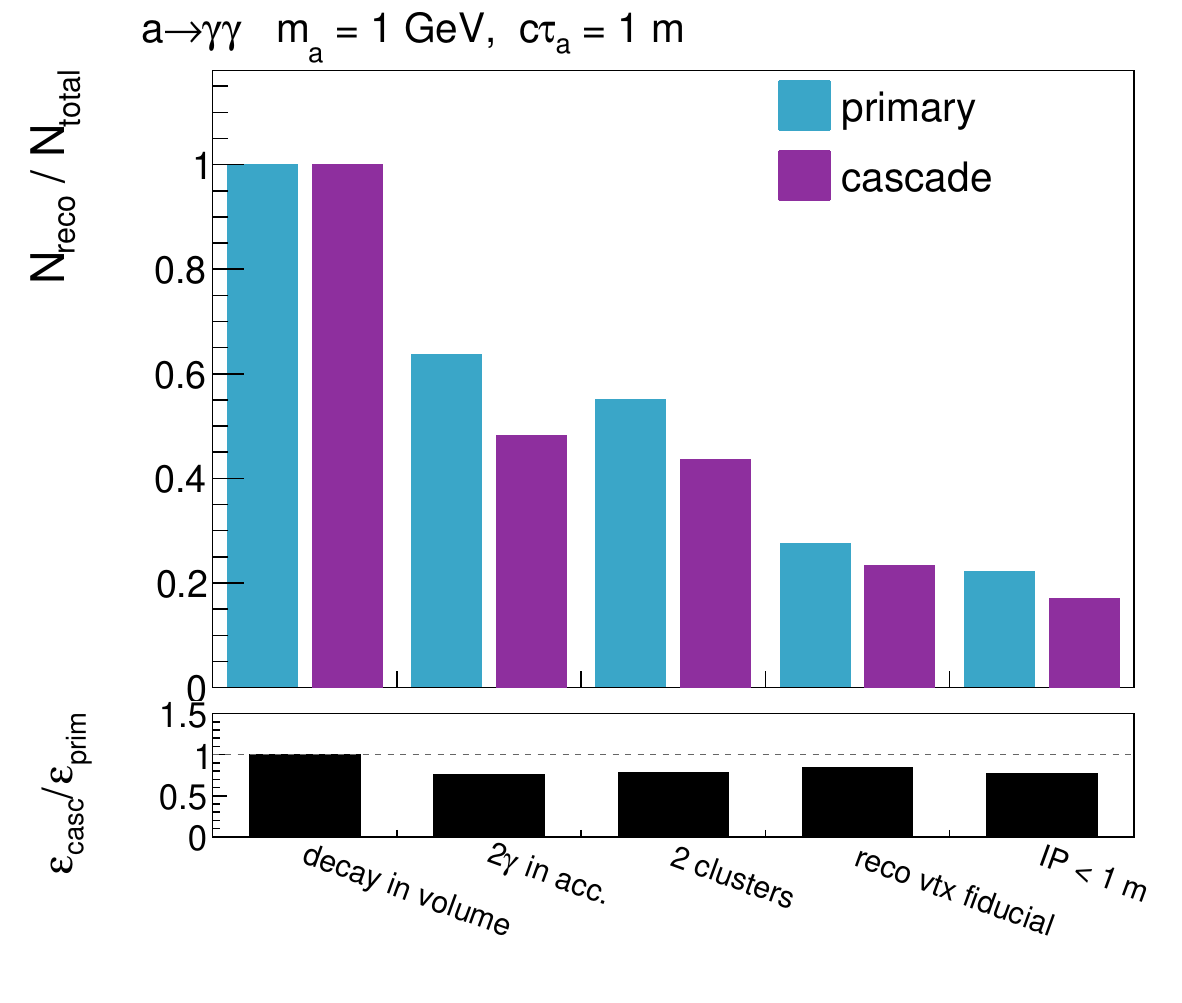}

\caption{Vertex, invariant-mass, and impact-parameter resolutions for a \SI{1}{\giga\electronvolt}, $c\tau_{a}=\SI{1}{\meter}$ ALP. The efficiencies after fiducial-volume selection cuts are also provided for successive selections: decays occurring in the decay volume, and both photons in acceptance are taken at the truth level. Events with two reconstructed clusters, events with a vertex in fiducial acceptance, and events with an impact parameter below \SI{1}{\meter} are taken at the reconstruction level. Each selection includes the previous conditions.}
\label{fig:benchmark}
\end{figure*}

To demonstrate the effects of adding the selection criteria, we consider the differential distribution of the decay events in the ALP energy $E_{a}$, which is obtained by integrating Eq.~\eqref{eq:Nevents-semi-analytic} in all quantities but $E_{a}$. We consider the primary and secondary ALPs for two reference masses, $m_{a} = 100\,\mev$ and $m_{a} = 500\,\mev$. We also consider the lifetime $c\tau_{a} = 300\,\m$, such that we are in the regime where the differential decay probability in Eq.~\eqref{eq:Nevents-semi-analytic} scales as $dP_{\rm decay}/dz \approx 1/(c\tau_{a}\sqrt{\gamma_{a}^{2}-1}\cos(\theta_{a}))$, and the differential number of events has a simple scaling with the ALP coupling, $dN_{\rm ev}/dE_{a} \propto g_{a}^{4}$. We focus on large lifetimes because the contribution of the cascade particles is most significant in this area of the parameter space: the small-lifetime regime prefers high-energy particles, which are naturally more populated by the primary production mechanisms.

The comparison is shown in Fig.~\ref{fig:energy-distributions-ALP}. After including the detector geometric acceptance, the distributions show a significant suppression of the event rate in the low-energy ALP tail. This is the effect of the minimal opening angle between the decay products, which prevents both photons from simultaneously flying to the detector even if the mother ALP points to it, as discussed in Sec.~\ref{ssec:existing-studies-limitations}. The effect is much more prominent for cascade events, as they primarily populate this region of the ALP kinematics.

Adding the energy cut drops the entire energy range $E_{a}< 2\,\gev$, and somewhat suppresses the event fraction for the range $2\,\gev < E_{a}\lesssim 10\,\gev$, because there is a non-negligible fraction of events where one of the photons is below the energy threshold.

The same daughter-level acceptance effect appears for HNLs produced in decays of secondary kaons. In this case, the relevant comparison is between primary HNLs, produced mostly by decays of $D$ mesons, and secondary HNLs from kaon decays. For illustration, Fig.~\ref{fig:energy-distribution-HNL} shows the differential event rate for HNLs mixing with $\nu_{e}$ and having mass $m_{N}=300\,\mev$.

Compared to the ALP case, the suppression of cascade HNLs is even stronger. Secondary kaons have a broad angular distribution, and the HNLs produced in their decays inherit this feature. Therefore, even before considering a dedicated detector-level reconstruction, the purely geometric requirement on the visible decay products already removes a large fraction of the cascade contribution.

For the mixing with $\nu_{\mu}$, all the conclusions qualitatively hold, except for slightly better geometric acceptance. Indeed, decays of charged-current HNL mixing with $\nu_{\mu}$ contain $\mu$ instead of $e$, meaning that the decay products have less momentum in the rest frame and hence less angular spread.

\subsubsection{Detector-level reconstruction}
\label{ssec:discussion-ALPs-detector-level}

Let us now turn to the detector-level reconstruction of the ALP signal. A representative working point, $m_a = 1\,\gev$ and $c\tau_a = 1\,\m$, is shown in Fig.~\ref{fig:benchmark} to illustrate the absolute scale of the vertex, invariant-mass, and impact-parameter resolutions. Here, however, the main question is comparative: whether the detector reconstructs the soft cascade population as efficiently as the harder primary one.

\begin{figure*}[t!]
    \includegraphics[width=\textwidth]{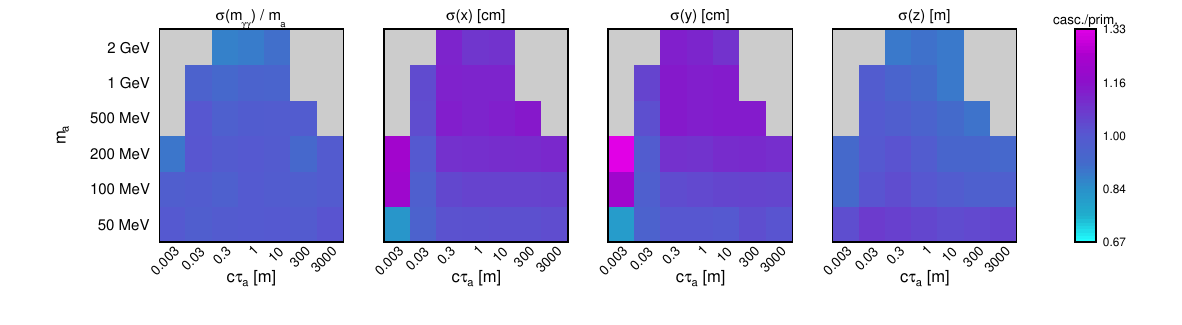}
    \caption{Ratios of the resolution in various reconstructed quantities between the events with cascade and primary ALPs, for different masses $m_{a}$ on the $y$ axis and $c\tau_{a}$ on the $x$ axis. From left to right, the panels show the invariant mass $m_{\gamma\gamma}$ and the $x, y, z$ coordinates of the ALP decay vertex. The $m_{a}=\SI{50}{\mega\electronvolt}$, $c\tau_{a}=\SI{1}{\meter}$ sample was interpolated between its same-mass neighbors.}
    \label{fig:det_sum_res}
\end{figure*}

This comparison is summarized in Fig.~\ref{fig:det_sum_res}, which shows the cascade-to-primary ratios of the reconstructed resolutions across the $(m_a,c\tau_a)$ plane. Because cascade ALPs are systematically softer than primary ones, their decay photons are lower in energy and wider in opening angle at fixed mass. This is the single kinematic characteristic behind the behavior of all panels.

For the transverse vertex coordinates, the cascade resolution is poorer across essentially the whole plane: softer photons give broader, lower-amplitude showers, whose PCA directions are noisier. As a result, the $\sigma(x)$ and $\sigma(y)$ ratios sit slightly above unity. The longitudinal resolution is instead governed by the di-photon opening angle through the point-of-the-closest-approach vertex, which is the worst-constrained direction for a two-photon system. The ratio here reflects the competition between the larger cascade opening angle, which improves the conditioning of the $z$ vertex, and the noisier per-cluster directions, and therefore stays closer to unity. Both primary and cascade resolutions degrade toward low ALP masses, where the minimum opening angle $\Delta\theta_{\min}\sim 2m_a/E_a$ collapses, and the photon pair becomes collinear. However, the degradation sets in earlier for the cascade population, which is why the lightest rows of Fig.~\ref{fig:det_sum_res} depart most strongly from unity. The dependence on $c\tau_a$ is the imprint of the surviving-energy spectrum: short lifetimes admit only the high-boost tail that reaches $z\gtrsim 33\,\m$, whose collimated photons inflate $\sigma_z$, while longer lifetimes admit a softer, wider-angle sample.

\subsection{Event-rate impact in the nominal SHiP setup}
\label{ssec:nominal-results}

We now combine the acceptance and reconstruction effects discussed above and evaluate the cascade contribution in the nominal SHiP setup as described in Sec.~\ref{sec:ship}. 

\begin{figure*}[t!]
    \includegraphics[width=\textwidth]{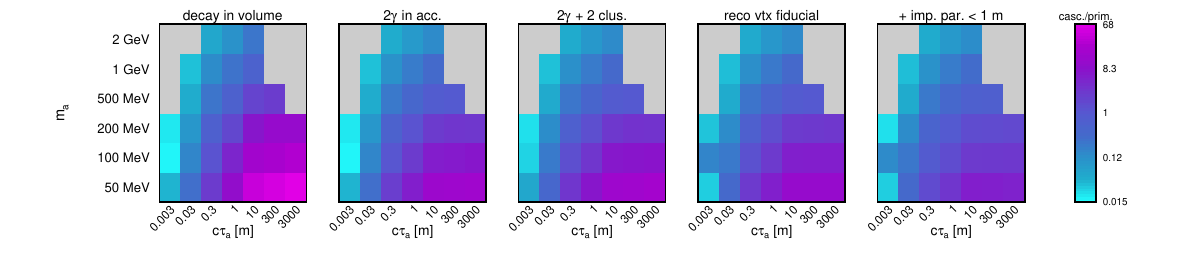}
    \caption{Observed ratios of the number of events between cascade and primary ALPs, for different ALP masses $m_{a}$ on the $y$ axis and lifetimes $c\tau_{a}$ on the $x$ axis, as obtained in the detector-level simulation. From left to right, the panels show the ratio of the number of events for ALPs decaying inside the decay volume; where both photons from the ALP decay are within the ECAL acceptance; where they form two separate clusters; where additionally the ALP decay vertex lies within the fiducial volume; and with the additional requirement on the transverse impact parameter, ${\rm IP}<1\,\m$. The first two panels correspond to the MC-truth results, whereas the rest are obtained using the detector simulation of the signal reconstruction. The $m_{a}=\SI{50}{\mega\electronvolt}$, $c\tau_{a}=\SI{1}{\meter}$ sample was interpolated between its same-mass neighbors instead of using the simulation data, due to an artifact in the latter.}
    \label{fig:det_sum_eff}
\end{figure*}

For ALPs, the detector-level cutflow is shown in Fig.~\ref{fig:det_sum_eff}. It follows the event sample from the decay-in-volume stage to the final reconstructed selection and, therefore, shows where the initial advantage of cascade production is lost throughout the reconstruction.

The efficiency comparison of Fig.~\ref{fig:det_sum_eff} shows a clear cutflow pattern. At the decay-in-volume stage, the reduced cascade boost raises the in-volume decay probability through the $1/\gamma_X$ scaling of Eq.~\eqref{eq:dPdecaydz}. Given this and also the discussion around Fig.~\ref{fig:overall-production-fluxes}, the cascade-to-primary ratio is largest toward large lifetimes and low masses. This advantage is then removed, step by step, by the requirements that make the event reconstructible.

The two-photon acceptance is the dominant suppressor. The wider opening angle of soft cascade decays makes it far less likely that both photons reach the $A_{\text{ECAL}}$ face. The two-cluster requirement removes a further low-mass fraction where the surviving pairs' increased collinearity degrades the clustering efficiency. Finally, the vertex and impact-parameter selections add an additional reconstruction-level suppression. The suppression coming from the impact parameter cut is again stronger for cascade events than for primary events (we comment on this in Sec.~\ref{ssec:selection-cuts}).

\begin{figure*}[t!]
    \centering
    \includegraphics[width=0.48\textwidth]{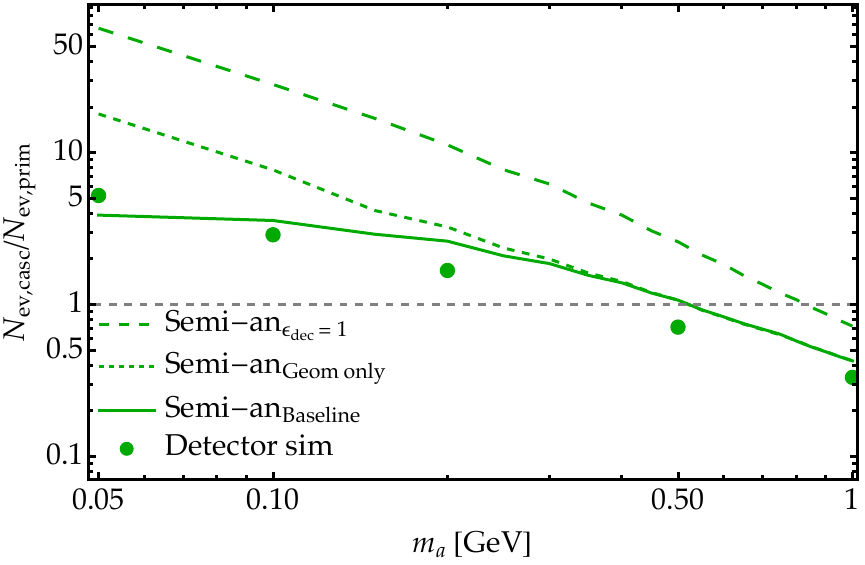}
    \includegraphics[width=0.48\textwidth]{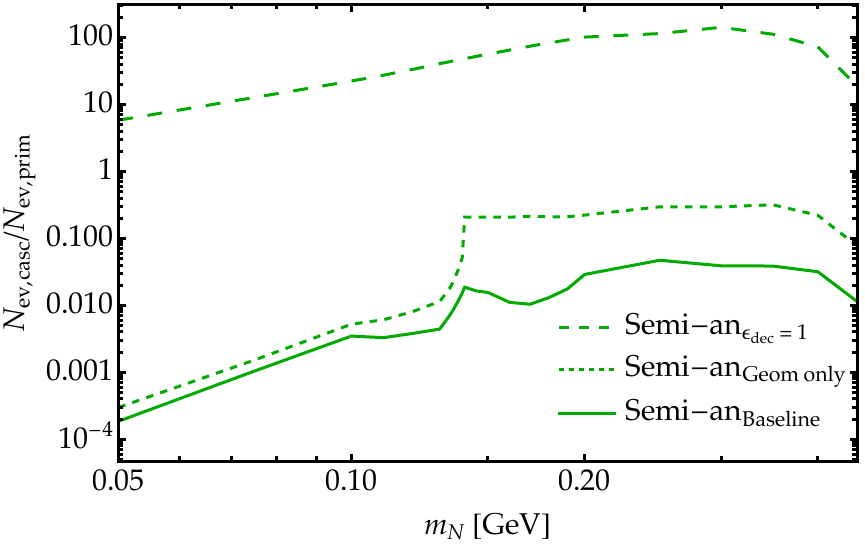}
    \caption{The ratio of the number of events with LLPs produced from cascade sources versus those having primary origin. Left panel: photophilic ALPs, for ALP lifetimes such that $c\tau_{a}\langle\gamma_{a}\rangle\gtrsim 100\,\m$. The continuous curves are obtained by the semi-analytic approach with Eq.~\eqref{eq:Nevents-semi-analytic}, and the meaning of the legends is the same as in Fig.~\ref{fig:energy-distributions-ALP}. The points come from the detector-level simulation described in Secs.~\ref{sec:signal-reconstruction} and~\ref{ssec:discussion-ALPs-detector-level}. Right panel: HNLs mixing with electron neutrinos. The meaning of the curves is the same, modulo the different sources of primary and cascade HNLs, as specified in the text. The non-monotonic behavior of the HNL ratio is caused by opening up various HNL decay modes at $m_{N}\simeq m_{\pi}$, such as $N\to \pi^{0}\nu$ and $N\to \pi + e$.}
    \label{fig:Nevents-ratio}
\end{figure*}

Let us now compare the integrated event rates from primary and cascade production. We use the simplified semi-analytic analysis for both ALPs and HNLs and the full detector-level simulation for ALPs. The quantity shown in Fig.~\ref{fig:Nevents-ratio} is
\begin{equation}
N_{\rm ev,cascade}/N_{\rm ev,primary}
\label{eq:Ncasc-to-Nprim}
\end{equation}
as a function of the LLP mass, in the regime where the decay length is large compared to the SHiP detector scale.

Let us start with the ALP result in the left panel of Fig.~\ref{fig:Nevents-ratio}. The ``$\epsilon_{\rm dec}=1$'' line shows a huge domination of cascade events over the primary ones. However, already requiring both photons to be within the acceptance of the ECAL, shown by the ``Geom only'' curve, suppresses the secondary yield by the relative factor of $\simeq 4$ compared to the $\epsilon_{\rm dec}=1$ line. Finally, assuming the baseline selection criteria, the ratio~\eqref{eq:Ncasc-to-Nprim} decreases further, especially in the domain of light ALPs with $m_{a}\lesssim 200\,\mev$. Overall, for a 50-MeV ALP, the ratio decreases from $\simeq 70$ to $\simeq 3$ when going from ``$\epsilon_{\rm dec}=1$'' to ``Baseline''.

On top of this, in the same panel, we show the results of the detector-level analysis. Given that it uses \textsc{EventCalc} for the signal generation, and \textsc{EventCalc} agrees well with \textsc{SensCalc}, the comparison between the ratios shows how realistic detector-level reconstruction differs from simple Monte Carlo truth cuts. The detector-level ratio sits close to the ``Baseline'' curve, and its dependence on the ALP mass follows a similar trend except for the domain $m_{a}\lesssim 100\,\mev$. There, the hard cut on the photon energy imposed in the semi-analytic calculation causes the plateau-like behavior of the cascade-to-primary ratio, whereas the detector simulation shows a monotonic (yet slow) growth. This indicates that the semi-analytic baseline selection captures the dominant suppression of the cascade ALP contribution, at least for the ratio of primary to cascade events.

The HNL result is shown in the right panel of Fig.~\ref{fig:Nevents-ratio}. The comparison shows that, already after including the purely geometric part of the decay-products acceptance, the number of events from cascade HNLs becomes subdominant to the primary source. Moreover, the kaon flux used here was obtained for the hybrid Molybdenum-Tungsten target, which has a larger kaon interaction length and hence gives more chances for the kaons to decay in flight. If switching to the actual Tungsten-only target, the flux of decaying kaons would become even more isotropic, suppressing the geometric acceptance for the HNLs and their decay products even further.

Given this, we do not perform a dedicated study of the reconstruction performance of low-energy HNL decay products. As discussed above, an additional complexity is that most of the visible HNL decay modes contain charged particles, and low-momentum species may be deflected out of the detector coverage by the spectrometer magnetic field. There is also the decay mode $N\to \pi^{0}\nu$, where the detectable final state is a two-photon system. However, since the cascade contribution is already subdominant after imposing geometric acceptance, we leave a detector-level study of such channels for future work.

\subsection{Sensitivity to the event-selection criteria}
\label{ssec:selection-cuts}

Let us now comment on the two selection requirements that most directly
affect the detector-level cascade ALP yield: the threshold on the energy
of the individual decay photons, Eq.~\eqref{eq:selection-ship-baseline-1},
and the impact-parameter requirement, Eq.~\eqref{eq:selection-ship-baseline-3}.

\begin{figure*}[t!]
    \centering
    \includegraphics[width=\linewidth]{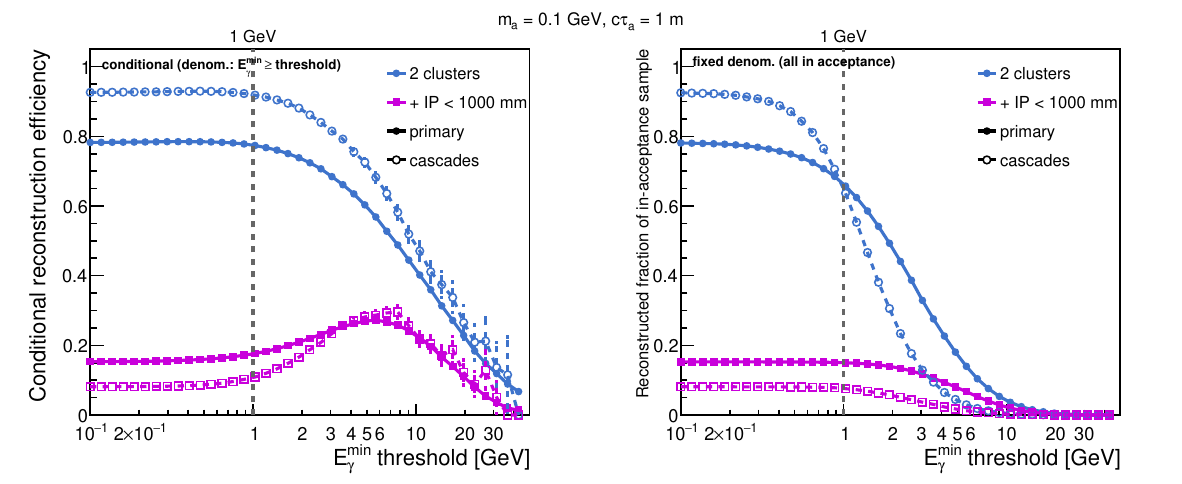}
\caption{Reconstruction performance for 100 000 $a\to\gamma\gamma$ decays as a function of a threshold $E_{\gamma}^{\rm min}$ imposed on the \emph{truth-level} energy of the \emph{softer} of the two decay photons, for the benchmark $m_{a}=100\,\mev$, $c\tau_{a}=1\,\m$, a low mass where cascade enhancement is significant. The mass is the same as in Fig.~\ref{fig:ip_cut}, which shows several lifetimes at this mass including $c\tau_{a}=1\,\m$, so that the two figures can be read together. Both panels are restricted to decays for which both photons lie within the ECAL acceptance, so that the geometric acceptance --- whose impact is shown separately in Figs.~\ref{fig:benchmark} and~\ref{fig:det_sum_eff} --- is removed and only the detector-level response is displayed. Left panel: conditional reconstruction efficiency; at each $E_{\gamma}^{\rm min}$ the numerator is the number of in-acceptance events above the threshold that are successfully reconstructed under the stated conditions (two reconstructed clusters, with or without the additional impact-parameter requirement), and the denominator is the number of in-acceptance events above the same threshold, so that the denominator moves with said threshold. Right panel: cumulative surviving fraction, i.e.\ the same numerator normalized to the fixed denominator of \emph{all} in-acceptance events; here ``cumulative'' refers to the fraction of the signal surviving a cut at $E_{\gamma}^{\rm min}$, so this panel measures directly how much signal is discarded by the nominal criterion~\eqref{eq:selection-ship-baseline-1}.}
\label{fig:egamma_cut}
\end{figure*}

We begin with the energy threshold. Fig.~\ref{fig:egamma_cut} quantifies the ECAL response to a threshold $E_{\gamma}^{\rm min}$ applied to the truth-level energy of the softer of the two decay photons; since Eq.~\eqref{eq:selection-ship-baseline-1} constrains each photon separately, an event survives the criterion if and only if its softer photon clears the threshold. Both panels are conditioned on the two photons being within the ECAL acceptance, so that the geometric suppression --- quantified in the cutflow of Fig.~\ref{fig:benchmark} and, in cascade-to-primary ratio, in the second panel of Fig.~\ref{fig:det_sum_eff} --- is removed. The left panel shows the conditional reconstruction efficiency, evaluated on the subsample passing the threshold: \ch{for the two-cluster requirement it is flat at $\simeq 0.78$ (primary) and $\simeq 0.93$ (cascade) below $\simeq 1\,\gev$, and decreases above it for both origins. The decrease reflects the fact that, at this low mass, raising the threshold selects progressively more boosted decays, whose photons are increasingly collimated and degrade clustering efficiency.} The right panel shows the same numerator normalized to the full in-acceptance sample, and therefore gives the fraction of the signal retained by a cut at a given energy.

Comparing the low-threshold plateau of the right panel with its value at $1\,\gev$ gives the signal recoverable by relaxing Eq.~\eqref{eq:selection-ship-baseline-1}. \ch{At the two-cluster stage the primary sample falls from $\simeq 0.78$ to $\simeq 0.66$ at $1\,\gev$, so the nominal threshold removes $\simeq 0.12$ of the in-acceptance signal: $\simeq 15\%$ of the low-threshold plateau, or equivalently a $\simeq 18\%$ increase on the signal retained at $1\,\gev$ if the criterion were fully relaxed. For cascade ALPs the corresponding drop is from $\simeq 0.93$ to $\simeq 0.64$, i.e.\ $\simeq 0.29$, which is $\simeq 31\%$ of the plateau or a gain of $\simeq 45\%$ on the retained signal.} Relaxing the threshold therefore recovers proportionally more cascade signal, as expected from their softer spectrum. Once the impact-parameter requirement is added, however, both curves are essentially flat across $1\,\gev$, so that lowering the threshold recovers almost nothing: the soft events admitted by a relaxed Eq.~\eqref{eq:selection-ship-baseline-1} are events that the impact-parameter requirement rejects in any case. The decrease of these curves at higher thresholds instead reflects the loss of genuinely reconstructible high-energy events and is not relevant to relaxing the criterion. Relaxing the energy threshold is therefore useful only in combination with a relaxed impact-parameter requirement, which we return to below.

Because the cascade spectrum is much softer
($\langle E_{\gamma,{\rm casc}}\rangle\approx 1\,\gev$ versus
$\langle E_{\gamma,{\rm prim}}\rangle\approx 16.6\,\gev$, recall
Sec.~\ref{ssec:SM-fluxes}), cascade events populate the low-threshold part of
these curves, whereas primary events sit close to the plateau. This spectral
effect is not, however, the whole story: at fixed softer-photon energy the
primary and cascade samples are not equivalent, and the comparison changes sign
between the two selection stages. At the two-cluster level the cascade curve lies above the primary one, \ch{by
$\simeq 0.15$ in absolute efficiency at low threshold, the two converging as the threshold increases}: at a given softer-photon energy the
cascade sample carries lower total ALP energy, so the decay opening angle is wider and the two showers are more readily resolved into separate clusters provided that both reach the calorimeter acceptance.
Adding the requirement ${\rm IP}<\SI{1000}{\milli\meter}$ reverses the ordering
below $\simeq 6\,\gev$ and reduces the cascade efficiency to roughly half the
primary one, since the decay direction is reconstructed from the cluster
positions and degrades both with softer photons and with the shorter lever arm
available for less boosted decays (recall
Sec.~\ref{ssec:existing-studies-limitations}). The suppression of the cascade
contribution after the full selection is therefore not of purely spectral
origin: a residual detector-level penalty survives at fixed photon energy, and
it enters through the directionality requirement rather than through cluster
reconstruction. The non-monotonic shape of the impact-parameter curves, which
peak near $5\text{--}7\,\gev$, follows from two competing trends visible in the
same panel: the fraction of reconstructed events passing the IP requirement grows with photon energy as the directional resolution improves, while the \ch{two-cluster efficiency itself falls above $\simeq 1\,\gev$, as
described above, because the increasingly collimated decays lead to degraded clustering efficiency}. The peak marks the point at which
the second effect overtakes the first.

\begin{figure}[t!]
    \centering
    \includegraphics[width=0.48\textwidth]{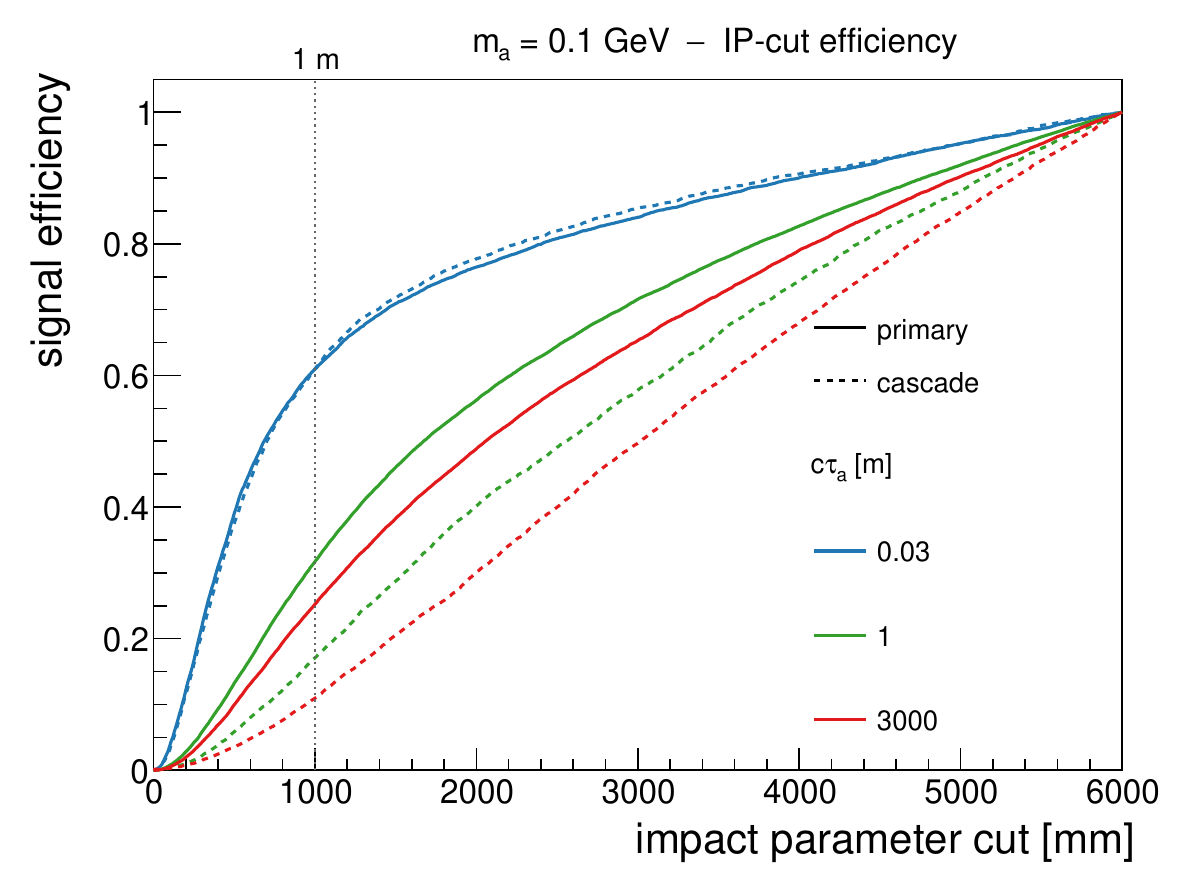}
    \caption{Relative signal efficiency as a function of the impact-parameter cut for the current reconstruction for several \SI{100}{\mega\electronvolt} ALP samples. The signal efficiency is observed to fall faster for cascade ALPs compared to primary ones in the larger-$c\tau_{a}$ cases, where the cascade enhancement is most significant.}
    \label{fig:ip_cut}
\end{figure}

Fig.~\ref{fig:ip_cut} shows the integrated counterpart of this effect, now as
a function of the impact-parameter cut itself and for the same benchmark mass
$m_{a}=\SI{100}{\mega\electronvolt}$, so that the two figures can be read
together. The requirement is especially detrimental to cascade events: because
cascade ALPs are softer and their reconstructed directions are less precise,
this cut removes relatively more cascade events than primary events.

If the impact parameter cut can be softened, or removed altogether for sufficiently clean two-photon final states, the cascade enhancement may become significantly larger. For example, for \SI{100}{\mega\electronvolt} ALPs with $c\tau_{a}>30\,\m$, a simple two-cluster requirement may lead to an enhancement of up to a factor of $\sim 8$. Whether this requirement can in practice be softened depends on the background composition and rejection power, and therefore requires a dedicated background study.

\begin{figure}[t!]
    \centering
    \includegraphics[width=\linewidth]{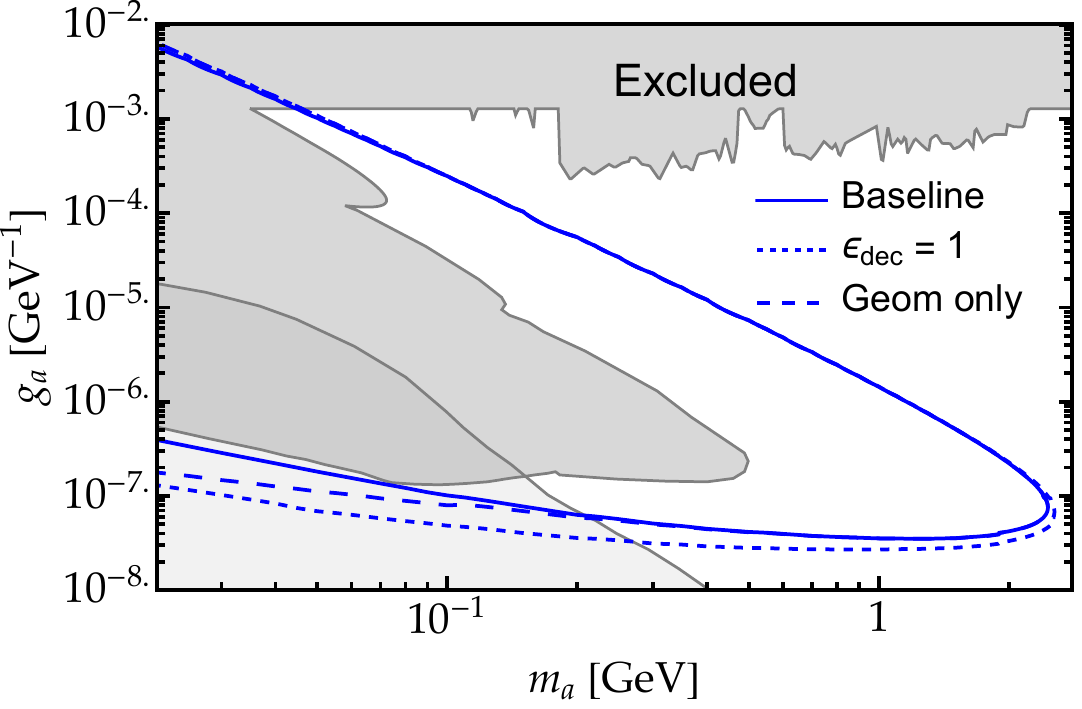}
    \caption{Sensitivity of the SHiP experiment to the ALPs obtained using the semi-analytic approach via Eq.~\eqref{eq:Nevents-semi-analytic}, as $N_{\rm ev}\geq 2.3$, in the plane ALP mass $m_{a}$-ALP coupling $g_{a}$. The meaning of the lines is provided in Tab.~\ref{tab:semi-analytic-event-classification}. The darker-gray and lighter-gray domains are constraints from laboratory searches and astrophysical observations, taken from Ref.~\cite{deBlas:2025gyz}. In the low-mass region, SHiP as a laboratory probe would cover the range of couplings complementarily to the astrophysical bounds.}
    \label{fig:ALP-sensitivity}
\end{figure}

To highlight the importance of the reconstruction of the low-energy events, we show in Fig.~\ref{fig:ALP-sensitivity} the sensitivity of SHiP to ALPs as obtained using the semi-analytic approach, under the requirements on decay products from Tab.~\ref{tab:semi-analytic-event-classification}. The difference between the $\epsilon_{\rm dec} = 1$ and ``Baseline'' lines at the lower bound of the sensitivity may reach a factor $\simeq 4$. Since the number of events scales as $N_{\rm ev}(g_{a})\propto g_{a}^{4}$, this would translate to a huge factor of $\lesssim 200$ enhancement in the event rates.

\section{Conclusions}
\label{sec:conclusions}

Cascade interactions inside the thick target of the SHiP experiment may produce long-lived particles (LLPs) whose decays can dominate the event rate inside the hidden-sector decay volume before detector-level requirements are imposed. In this paper, we have quantified the contribution of these cascade-induced signal sources relative to primary production mechanisms, focusing on the regime in which the produced LLPs are relatively soft, with typical energies $E_{\rm LLP}\lesssim \text{few GeV}$.

We considered two representative case studies: photophilic ALPs, discussed in Sec.~\ref{ssec:ALP-pheno}, and HNLs, discussed in Sec.~\ref{ssec:HNL-pheno}. These particles can be produced from cascade photons and secondary kaons, respectively, as described in Sec.~\ref{ssec:SM-fluxes}. Extending previous studies, we evaluated how the event rate is affected by two detector-level effects: first, the geometric acceptance of the detector, which requires the LLP decay products to remain within the SHiP detector system; and second, the reconstruction efficiency for low-energy decay products.

For the nominal SHiP setup, our main results are summarized in Figs.~\ref{fig:det_sum_eff} and~\ref{fig:Nevents-ratio}. For light LLPs, cascade production can significantly exceed the primary yield before imposing requirements on the decay products. However, the geometric acceptance of the detector already substantially suppresses this cascade-driven enhancement. For photophilic ALPs, the inclusion of realistic event reconstruction further reduces the effect, leaving only a moderate enhancement for the lightest masses. For HNLs from secondary kaons, the cascade contribution becomes subdominant already after imposing the geometric acceptance of the visible decay products.

At the same time, our results indicate that the SHiP search program could benefit from improving the detectability of such low-energy signals. One possibility is to enlarge the effective angular coverage of the electromagnetic calorimeter and to improve the background-rejection strategy, which may allow the transverse impact-parameter requirement to be relaxed. As shown in Fig.~\ref{fig:ip_cut}, this requirement removes cascade ALP events more efficiently than primary ones, especially in the long-lifetime region where the cascade enhancement is most relevant.

Another possibility is to consider more usage of active decay-volume subdetectors, where the decay products could be reconstructed closer to the LLP decay vertex. A straightforward option such as SND@SHiP does not appear well-suited for this purpose, given its small size and the intense muon background. An existing option would be the use of the SHiP Surrounding Background Tagger \cite{Brignoli:2025spc} to attempt reconstruction of LLP reaching the decay vessel but whose final states fail to (all) reach the downstream detectors. This would imply leveraging the detector's full coverage of the decay vessel and $\mathcal{O}(\si{\centi\meter})$ position resolution to reconstruct otherwise incomplete or fully missing events. The detector has poor efficiency for photon reconstruction but may allow to recover part of the cascade enhancement in the lower mass region.
A different possibility is an LAr-based setup placed at the end of the SHiP detector, as discussed in Ref.~\cite{Ferrillo:2023hhg}. A robust assessment of such an option would require a dedicated study of signal reconstruction and backgrounds, which we leave for future work.

This makes the present study particularly timely. By quantifying the event-rate suppression induced by realistic detector acceptance and reconstruction, we show that reliable sensitivity estimates for cascade-induced LLP signals require detector effects to be included up to the reconstruction level. At the same time, since SHiP is still in the TDR phase, there remains a concrete opportunity to optimize the setup and recover part of the cascade-induced event rate.

\section*{Acknowledgements}
The authors thank Yotam Soreq for useful discussions, Dmitry Gorbunov for discussions on the kaon production inside a thick SHiP target, and Nikita Blinov, Andrei Golutvin, Anne-Marie Magnan, and Ryan Plestid for useful comments on the manuscript. Matei Climescu has received support from the FWO, under grant no. 12A4O26N (Belgium).

\section*{Conflict of Interest Statement}
Some of the authors (MC, MO) are also members of the SHiP collaboration. The present manuscript solely reflects the authors' views and not those of the SHiP collaboration.

\bibliography{bib.bib}

\clearpage

\onecolumngrid

\appendix

\section*{Supplemental material}

\section{Comparison with Ref.~\cite{Patrone:2025fwk}}
\label{app:cross-check-ALPETITE}

To cross-check our calculation of the ALP flux, we have compared our results to Fig. 3 of Ref.~\cite{Patrone:2025fwk}. For these purposes, within \textsc{EventCalc}, we have obtained the ALP events energy distribution $dN_{\rm ev}/dE_{a}$ for mass $m_{a} = 100\,\mev$ and $c\tau_{a} = 300\,\m$, assuming no cuts on decay products, but imposing the cut $\theta_{a}<0.029\,{\rm rad}$ on the decaying ALP (this is the simplified analog of the decay products acceptance used in Ref.~\cite{Patrone:2025fwk}); the impact of this cut is quite low, because the ALPs are strongly collimated. Next, we have rescaled Fig.~3 of Ref.~\cite{Patrone:2025fwk} by $N_{\rm PoT}\times f_{c\tau_{a} = 300\,\m}^{-4} \approx 3.6\times 10^{5}\,\gev^{-4}$, where $f_{c\tau_{a} = 300\,\m}^{-1} \approx 1.6\times 10^{-4}\,\gev^{-1}$. Finally, we have rescaled it by the factor $(1-\exp(-1))^{-1}$ to account for a conservative assumption that only the first proton interaction length is included within the target when obtaining the primary photon flux.

\begin{figure}[t!]
    \centering
    \includegraphics[width=\linewidth]{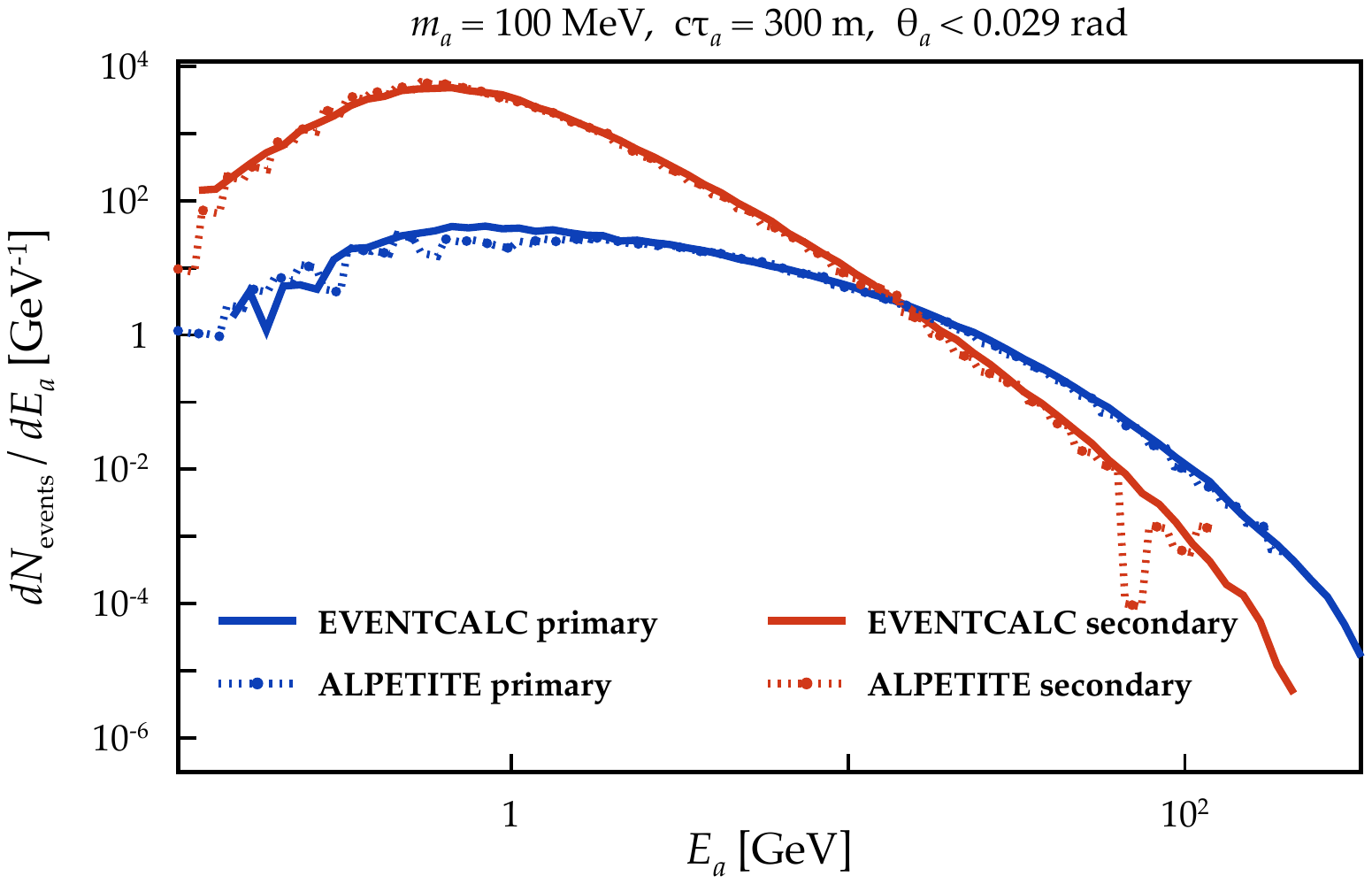}
    \caption{The energy distribution of the number of events with ALPs having mass $m_{a} = 100\,\mev$ and lifetime $c\tau_{a}=300\,\m$, assuming no cuts on decay products but requiring the ALP trajectory to satisfy $\theta_{a}<0.029\,\text{rad}$, as obtained by \textsc{EventCalc} (solid) and \textsc{ALPETITE} (dashed), for the primary and secondary event sources.}
    \label{fig:comparison-vs-ALPETITE}
\end{figure}

The resulting comparison is shown in Fig.~\ref{fig:comparison-vs-ALPETITE}. The results are in excellent agreement except for the domain $E_{a}\lesssim 1\,\gev$. The small discrepancy there is likely due to a combination of a slightly different primary photon flux used in the calculations and also different target materials assumed in the calculation (Molybdenum vs Tungsten). The agreement confirms the mutual agreement in calculating the ALP number of events -- from computing the photon flux to accounting for decay volume geometry when sampling ALP decay events.

\end{document}